\documentclass[aps,prf,onecolumn,superscriptaddress,groupedaddress,longbibliography]{revtex4}

\usepackage{bm}
\usepackage{graphicx}
\usepackage{url}
\usepackage{subfigure}
\usepackage{amsmath}
\usepackage{amssymb}
\usepackage{color}
\usepackage[normalem]{ulem}

\usepackage[a4paper,left=1.5cm,right=1.3cm,top=3.0cm,bottom=2cm]{geometry}

\renewcommand{\Pr}   {\ifmmode \mathrm{Pr}  \else $\mathrm{Pr}$ \fi} 
\newcommand{\Ra}     {\ifmmode \mathrm{Ra}  \else $\mathrm{Ra}$ \fi} 
\newcommand{\Nu}     {\ifmmode \mathrm{Nu}  \else $\mathrm{Nu}$ \fi} 
\renewcommand{\Re}   {\ifmmode \mathrm{Re}  \else $\mathrm{Re}$ \fi}

\usepackage{comment}

\begin{document}

\title{Strongly superadiabatic and stratified limits of compressible convection}

\author{John Panickacheril John}
\affiliation{Institut f\"ur Thermo- und Fluiddynamik, Technische Universit\"at Ilmenau, D-98684 Ilmenau, Germany}

\author{J\"org Schumacher}
\affiliation{Institut f\"ur Thermo- und Fluiddynamik, Technische Universit\"at Ilmenau, D-98684 Ilmenau, Germany}
\affiliation{Tandon School of Engineering, New York University, New York City, NY 11201, USA}
\email{joerg.schumacher@tu-ilmenau.de}

\date{\today}

\begin{abstract}
Fully compressible turbulent convection beyond the Oberbeck-Boussinesq limit and anelastic regime is studied in three-dimensional numerical simulations. Superadiabaticity $\epsilon$ and dissipation number $D$, which measures the strength of stratification of adiabatic equilibria, cause two limits of compressible convection -- nearly top-down-symmetric, strongly superadiabatic and highly top-down-asymmetric, strongly stratified convection. Highest turbulent Mach numbers $M_t$ follow for a symmetric blend of these two limits which we term the fully compressible case. Particularly, the strongly stratified convection case leads to a fluctuation-reduced top layer in the convection zone, a strongly reduced global heat transfer, and differing boundary layer dynamics between top and bottom. We detect this asymmetry for growing dissipation number $D$ also in the phase plane which is spanned by the turbulent Mach number $M_t$ and the dilatation parameter $\delta$ which relates the dilatational velocity fluctuations to the solenoidal ones. A detailed analysis of the different transport currents in the fully compressible energy budget relates the low-$D$ convection cases to the standard definition of the dimensionless Nusselt number in the Oberbeck-Boussinesq limit. 
\end{abstract}

\maketitle
\section{Introduction}
Buoyancy-driven turbulence deviates in most applications significantly from the thermal convection paradigm of a Rayleigh-Bénard layer heated from below and cooled from above \cite{Kadanoff2001}. The latter is characterized by small deviations from an (adiabatic) equilibrium state, incompressibility, and constant material properties thus leading to a perfect statistical top-down symmetry, termed the Oberbeck-Boussinesq (OB) limit \cite{Siggia1994,Ahlers2009,Chilla2012,Verma2018}. Non-Oberbeck-Boussinesq (NOB) effects are caused on the one hand by additional physical processes, such as phase changes in the atmosphere \cite{Stevens2005,Pauluis2011} which generate buoyancy reversals \cite{Mellado2017}, or due to strong dependencies of material properties and density on temperature and pressure \cite{Urban2012,Couston2017,Toppaladoddi2018,Pandey2021,Pandey2021a}, e.g., in technical fluids for nuclear engineering \cite{Pioro2016}. On the other hand, they are also caused by compressibility \cite{Froehlich1992}, such as in giant gas planets \cite{Jones2009,Gastine2012} or  the coupling of ocean mesoscale eddies to the surface wind stress fields of the lower atmosphere \cite{Emmanuel1986,Renault2017}. The corresponding  numerical models typically involve the anelastic (AE) limit of the compressible flow case representing  a small excess from the adiabatic equilibrium \cite{Lantz1999}. Other examples require a fully compressible treatment. Strong radiative cooling drives solar convection from the surface causing downflows with characteristic velocity fluctuations reaching the speed of sound, $u^{\prime} \sim c$ \cite{Stein2012,Rincon2018,Brandenburg2016,Schumacher2020}.  For supersonic convection, see e.g. refs. \cite{Cattaneo1990,Malagoli1990}, and for forced interstellar gas cloud turbulence with the velocity fluctuations up to $u^{\prime}\lesssim 5 c$, see ref. \cite{Federath2021}. 

In this work, we investigate fully compressible turbulent convection solely driven by buoyancy forces. The flow is characterized by the superadiabaticity, $\epsilon$, and a measure of the stratification of density or temperature, the dissipation number $D$. Both parameters are bounded and form a triangular $\epsilon$--$D$ parameter plane. Buoyancy-driven compressible convection can thus proceed in two limits -- the {\em strongly superadiabatic} (SAC) and the {\em strongly stratified convection} (SSC) -- which lead to subsonic root mean square (rms) velocity amplitudes $u^{\prime}$. Their mixture, which we will term {\em fully compressible convection} (FCC), is shown to result in the largest turbulent Mach numbers, $M_t=u^{\prime}/c$. A first objective of the present work is consequently a systematic exploration of the $\epsilon$--$D$ parameter plane. 

Variable material properties such as the temperature and pressure dependence of dynamic viscosity $\mu(T,p)$ and thermal conductivity $k(T,p)$ can also lead to non-Oberbeck-Boussinesq (NOB) behavior \cite{Chilla2012}. This is relevant in many astrophysical and geophysical systems, where we have gaseous systems with a complicated equation of state involving non-trivial dependencies of material properties on thermodynamic state variables, $T$ and $p$. They also differ from the usually assumed Sutherland law for ideal gases \cite{Sutherland1893}. In reality, these variable fluid properties can interact with compressibility effects which complicates a systematic analysis of the convection regimes. This is outside the scope of the present study. Here, we try to disentangle and isolate {\em genuine compressibility effects} from  other possible routes to NOB convection \cite{Horn2013}, by keeping constant material parameters, $\mu_0$ and $k_0$.

We also show that not every NOB flow will generate highly asymmetric mean vertical profiles of central turbulence quantities across the layer even though the detaching thermal plumes are found to be asymmetric in all cases -- thinner plumes detaching from the top in comparison to the ones from the bottom. We will use the Kullback-Leibler (KL) divergence \cite{Goodfellow2016} between the {\em local} boundary layer thickness distributions at top and bottom to access this asymmetry for all cases \cite{Scheel2014}. Finally, we analyse in detail the energy equation budget, such that we can compare the different currents with the diffusive and convective heat currents from Rayleigh-B\'{e}nard convection case which is the OB limit with $\epsilon \to 0$ and subsequently $D \to 0$.

One motivation for the present study of the SSC is the interior of the Sun. In the upper layer, the turbulent Mach number is of ${\cal O}(1)$ or partly even slightly larger at the surface to the coronal vacuum, possible due to a moderate superadiabaticity and extreme drops of temperature and density. The strong stratification, particularly in the upper convection zone, is in line with anomalously weak velocity fluctuations, as observed by helioseismology measurements \cite{Hanasoge2012} for solar radii $r\gtrsim 0.92 R_{\odot}$. By exploring the extreme limits of turbulent compressible convection, as done here in a much simpler setting, our results might suggest new pathways to better explain these observations, despite being orders of magnitude away from reality in terms of Rayleigh and Prandtl numbers. We will come back to this point in the final discussion of the work.

\section{Governing equations}
Our study is based on three-dimensional direct numerical simulations (DNS) of the fully compressible differential balance equations for mass, momentum, and internal energy densities, $\rho$, $\rho u_i$, and $\rho e$, by a compact finite difference scheme \cite{Lele1992,JDJFM2016,Kumari2020} which is appropriate for subsonic flows \cite{Pirozzoli2011}. The fully compressible Navier-Stokes  equations along with the buoyancy term are given by,
\begin{subequations}
    
\begin{equation}
 \frac{\partial \rho} {\partial t} + \frac{\partial \left(\rho u_{i}\right)}{\partial x_{i}} = 0
\label{eq:mass}
\end{equation}
\begin{equation}
\frac{\partial\left(\rho u_{i}\right)}{\partial t} + \frac{\partial \left(\rho u_{i} u_{j} \right)}{\partial x_{j}}  = -\frac{\partial p }{ \partial x_{i}} + \frac{\partial \sigma_{ij}}{\partial x_{j}} - \rho g \delta_{i,3} %\colr{\mathbf{f_{i}}}
\label{eq:mom}
\end{equation}
\begin{equation}
\frac{\partial \left(\rho e\right)}{\partial t} + \frac{\partial \left(\rho e u_{j} \right)}{\partial x_{j}}  = -p \frac{\partial u_{i} }{\partial x_{i}} +
\frac{\partial}{\partial x_{i}}\left(k \frac{\partial T}{\partial x_{i}}\right)
+ \sigma_{ij} S_{ij} % - \colr{\mathbf{\wedge}}
\label{eq:ener}
\end{equation}
\begin{equation}
p = \rho R T \quad\textrm{ where }\quad R= C_{p} -C_{v} \,, 
\end{equation}
\end{subequations}
with $i,j=x,y,z$. These equations correspond to mass, momentum and energy conservation laws along with the ideal gas equation of state.  Here, $\rho$, $u_{i}$, $p$, $\rho e$, $T$ are the density, components of the velocity field, pressure, internal energy density and temperature, respectively. The specific internal energy is defined as $e= C_{v} T$. $R$ is the gas constant. We used the Einstein summation convention. The viscous stress tensor, $\sigma_{ij}$, and strain rate tensor, $S_{ij}$, are defined as 
\begin{equation}
\sigma_{ij} = \mu \left(\frac{\partial u_{i}}{\partial x_{j}} + \frac{\partial u_{j}}{\partial x_{i}}
- \frac{2}{3} \delta_{ij} \frac{\partial u_{k}} {\partial x_{k}}\right) 
\quad \mbox{and} \quad  S_{ij} =  \frac{1}{2}\left(\frac{\partial u_{i}}{\partial x_{j}} + \frac{\partial u_{j}}{\partial x_{i}} \right)\,,
\end{equation}
respectively. The dynamic viscosity is assumed to be constant in these simulations, $\mu=\mu_0$. The thermal conductivity, $k$ is related to the viscosity through the Prandtl number, $k_0= \mu_0 C_{p}/Pr$. In all our simulations, the Prandtl number is assumed to be $Pr= 0.7$. Here, $g$ is the acceleration due to gravity, $C_{p}$ and   $C_{v}$ correspond to specific heat at constant pressure and volume, respectively. Their ratio, the adiabatic coefficient, $\gamma= C_{p}/C_{v}= 1.4$ corresponds to a diatomic gas.

We have modified the code used in \cite{Baranwal2022} for simulating channel flows. The space is discretized using horizontally uniform grids in the $x$ and $y$ direction along with periodic boundary conditions.   Along the wall normal direction, non-uniform grid is used with clustering near the walls by using a hyperbolic tangent stretching function. The spatial derivatives are calculated using $6^{th}$-order compact finite differences for all points except near the walls. Fourth- and third-order compact schemes are used at the last two point grid points near the wall respectively. No-slip boundary conditions are used for the  walls.  The boundary condition for pressure is evaluated using the $z$-direction Navier-Stokes equations at the wall 
\begin{equation}
\frac{\partial p }{ \partial z} =  \frac{\partial \sigma_{iz}}{\partial x_{i}} - \rho g\,.
\end{equation}

The variables are advanced in time using low storage third-order Runge-Kutta scheme. A Courant-Friedrichs-Lewy (CFL) number of $0.5$ is used for all our direct numerical simulations.  All the statistics are taken after the flow has reached steady state.

\begin{table}
\renewcommand{\arraystretch}{1.2}
\begin{center}
\begin{tabular}{ccccccccccccccc} 
\hline\hline
No.  & Identifier & $D$ &  $\epsilon$ & $T_{\rm bot}/T_{\rm top} $  &   $ \bar\rho_{\rm bot}/\bar\rho_{\rm top}  $    &        $M^{\rm max}_{t}$ & $Ra$ & $Pr$ & $\gamma$ & $N_x \times N_y \times N_z$ & $Re$ &  $Nu$ &  \# $\lambda^{\rm bot}_{T}$      & \# $\lambda^{\rm top}_{T}$       \\ 
\hline
$1_{\rm RBC}$ & RBC & $0$ & $0$ & -- & --   & $0$ & $10^5$ & $0.7$ & $1.4$  & $256,000\times 5^3$ & 94 $\pm$ 0.3 & 4.3 $\pm$ 0.02 & 66 & 66\\ 

$1$ & OB & $0.1$ &  $0.1$ & $1.25$ &  $1.3$  &$0.12$    &$10^5$  & 0.72 & 1.4 &  $256 \times 256 \times 128$ & 117 $\pm$ 0.8  &4.0 $\pm  0.03$ & $28$ & $29$  \\

$2$ & SAC  & $0.1$ & $0.8$ &  $10$ & $1.3$ &$0.78$   &$10^5$ & 0.72 & 1.4 & $256 \times 256 \times 128$ & 175 $\pm$ 2.1 & 5.1 $\pm$ 0.03 & $27$ & $23$\\
 
$3$ & FCC & $0.5$ & $0.45$ & $20$ & $5.7$ &$1.12$  &$10^5$ & 0.72 & 1.4 & $256 \times 256 \times 128$ & 123 $\pm$ 0.3 & 4.0 $\pm$ 0.06 & $21$ & $34$\\

$4$ & SSC & $0.8$ & $0.1$ &  $10$ & $55.9$  &$0.39$  &$10^5$  & 0.72 & 1.4 & $256 \times 256 \times 128$ & 102 $\pm$ 1.5 & 2.8 $\pm$ 0.05 & $18$ & $48$\\
 
$5_{\rm RBC}$ & RBC & $0$ & $0$ & -- & -- &$0$ &   $10^6$ & $0.7$ & $1.4$  & $256,000\times 7^3$ & 296 $\pm$ 2.8& 8.1 $\pm$ 0.06 & 58 & 58\\ 

$5$ & OB & $0.1$ &  $0.1$ & $1.25$ & $1.3$ &$0.14$    &$10^6$ & 0.72 & 1.4 & $512 \times 512 \times 256$ & 424 $\pm$ 3.9 & 7.9 $\pm$ 0.04 & $35$ & $38$ \\

$6$ & SAC & $0.1$  & $0.8$ & $10$ & $1.3$ &$0.84$     & $10^6$ & 0.72 & 1.4 & $512 \times 512 \times 256$ & 582 $\pm$ 2.6 & 9.9 $\pm 0.01$ & $33$ & $29$\\

$7$ & FCC & $0.5$ & $0.45$ & $20$ & $5.7$ &$1.32$    &$10^6$ & 0.72 & 1.4 & $512 \times 512 \times 256$ &437 $\pm$ 3.9 & 8.0 $\pm$ 0.04 & $26$ & $44$\\

8 & SSC & $0.8$ & $0.1$ & $10$ & $55.9$ &$0.61$&  $10^6$   & 0.72 & 1.4 & $512 \times 512 \times 256$ &351 $\pm$ 3.1 & 4.3 $\pm$ 0.04 & $24$ & $74$ \\
\hline\hline 
\end{tabular}
\end{center}
\caption{
\label{tab:simusupp} 
List of 8 compressible direct numerical simulations (DNS) along with the data from 2 incompressible RBC cases. The identifier, dissipation number $D$, superadiabaticity $\epsilon$, ratio of temperature and density (adiabatic) at the bottom and top, maximum turbulent Mach number $M_t^{\rm max}$,  Rayleigh number $Ra$, Prandtl number $Pr$, adiabatic coefficient $\gamma$, grid dimensions, Reynolds number $Re$, Nusselt number $Nu$, and number of horizontal grid planes inside the thermal boundary layers at the bottom and top are listed. All simulations have an aspect ratio of $\Gamma=L/H=4$. The error bars for $Re$ and $Nu$ are obtained from the difference of the corresponding values of the two halves of the snapshot series. We add two Rayleigh-B\'{e}nard convection runs (subscript RBC) at the corresponding Rayleigh numbers which have been obtained by spectral element simulations \cite{Scheel2013} at $(D,\epsilon)=(0,0)$ as references. Their grid resolution is given by the number of spectral elements times the number of collocation points on each spectral element in all three space directions.}
\end{table}

\section{Compressible Convection Regimes}
 The adiabatic equilibrium, for which the compressible convection layer is at rest, is given by $\bar{T}(z)=\bar{T}_{\rm bot}(1-Dz/H)$. The reference temperature is the adiabatic value at the bottom of the layer, $T_{\rm bot}=\bar{T}_{\rm bot}$. The first new dimensionless parameter $D$ relates the dry adiabatic lapse rate $g/C_p$ to a characteristic temperature drop across the convection layer of height $H$ and thus measures the strength of stratification. It is termed the {\em dissipation number} \cite{Verhoeven2015} and given by 
 \begin{equation}
D=\frac{gH}{C_p T_{\rm bot}}=\frac{T_{\rm bot}-\bar{T}_{\rm top}}{T_{\rm bot}}\,.
\label{def1a}
\end{equation}
Bars denote plane-time averages of adiabatic profiles. The actual value at the top differs with  $T_{\rm top} < \bar{T}_{\rm top} (<T_{\rm bot})$ for convection to occur. The second new dimensionless parameter in compressible convection is the {\em superadiabaticity}, $\epsilon$ \cite{Verhoeven2015} which is defined as 
\begin{equation}
\epsilon=\frac{\bar{T}_{\rm top}-T_{\rm top}}{T_{\rm bot}}\,.
\label{def1b}
\end{equation} 
Note that $\epsilon \rightarrow 0$ corresponds to the anelastic limit, where the high frequency acoustic motions are filtered out of the system. This is possible due to the scale separation between convective and acoustic motions at low $\epsilon$. However at a finite $\epsilon$, acoustic motions can interact with the convective physics which can lead to NOB behavior. 

\begin{figure}[t]
\centering
\includegraphics[width=0.85\linewidth]{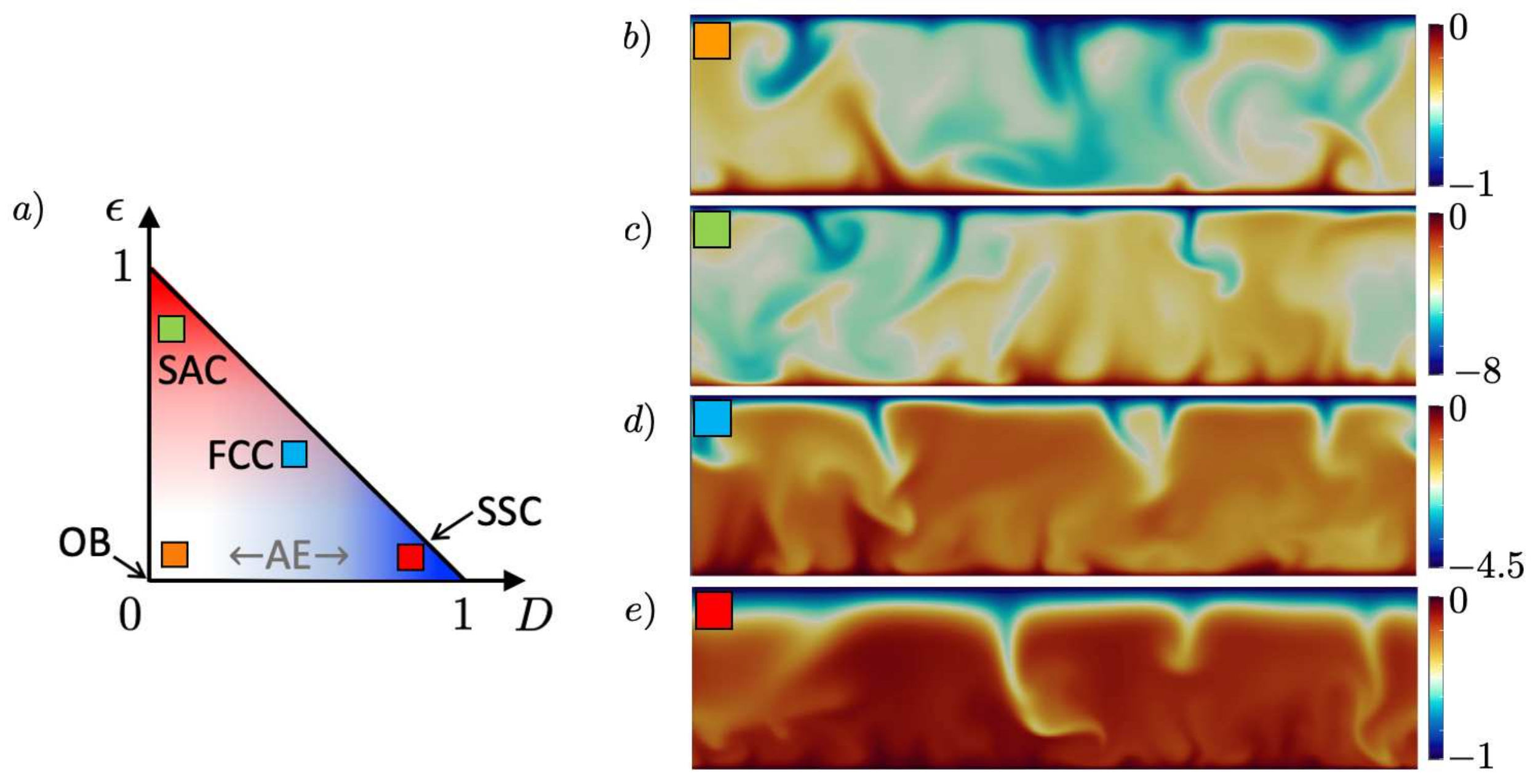}
\caption{Different regimes of compressible turbulent convection. (a) Parameter plane spanned by dissipation number $D$ and superadiabaticity $\epsilon$. The exact Oberbeck-Boussinesq limit (OB) at the origin and the anelastic (AE) regime are indicated together with the 4 different simulation cases. (b--e) Vertical slice cuts through instantaneous temperature $T_{\rm sa}$ at $(\epsilon, D)=(0.1, 0.1)$ near the OB limit in (b), in highly superadiabatic convection (SAC) at $(0.8, 0.1)$ in (c), in fully compressible convection at  $(0.45, 0.5)$ in (d), and in strongly stratified convection (SSC) at $(0.1, 0.8)$ in (e). The Rayleigh number is $Ra\simeq 10^6$, the Prandtl number $Pr=0.7$ in all cases.}
\label{Fig1}
\end{figure}

For incompressible Rayleigh B\'enard convection, the limits $\epsilon\to 0$ and subsequently $D \to 0$ are taken \cite{Verhoeven2015}. Thus NOB effects can be observed for finite magnitudes of both, $\epsilon$ and $D$. From eqns. \eqref{def1a} and \eqref{def1b}, it is also clear that both dimensionless parameters satisfy $0 \le \epsilon, D \le 1$. Furthermore, we get $\epsilon + D= \Delta T/T_{\rm bot} = \left(T_{\rm bot} - T_{\rm top} \right)/T_{\rm bot}$. Thus, it follows immediately that $\epsilon \le 1-D$ which restricts the $\epsilon$--$D$ parameter plane to a triangle as shown in Fig.~\ref{Fig1}(a). In this paper, we are interested in analysing the limits of compressible convection in terms of both the compressibility parameters. They are found in the corners of the parameter plane. When defining the free-fall velocity by $U_f=\sqrt{g\epsilon H}$ and the speed of sound by $c_s=\sqrt{\gamma R T_{\rm bot}}$, a {\em free-fall Mach number} $M_f$ can be defined 
\cite{Verhoeven2015}. The parameter $M_f$ is given by 
%------------------------------------------------------------------------
\begin{equation}
M_f=\frac{U_f}{c_s} = \sqrt{\frac{\epsilon D}{\gamma-1}} \le 1 \quad \mbox{for} \quad \epsilon, D\in [0,1]\,.
\label{def2}
\end{equation}          
%--------------------------------------------------------------------------
Equation \eqref{def2} shows that $M_f$ is bounded from above by 1. For this bound we use $\gamma=7/5=1.4$ of an ideal gas and $\epsilon\le 1-D$. The maximum $M_f^{\ast} = \sqrt{5/8}\approx 0.79$ is obtained for a symmetric blend, $D=\epsilon=1/2$. We can now identify different regimes of compressible thermal convection:
\begin{enumerate}
   \item {\em Oberbeck-Boussinesq-like convection} exists for $\epsilon\ll 1$ and $D\ll 1$ close to the exact OB Rayleigh-B\'{e}rnard convection case at $(\epsilon,D)=(0,0)$. This regime will be denoted for simplicity to as OB or OB-like in this work. In this particular study, we take $\epsilon = D= 0.1$.
   \item {\em Strongly stratified convection} exists for $\epsilon\ll 1$ and $D\to 1$ and will be denoted to as SSC. It is an extreme case of the AE limit that requires $\epsilon\ll 1$ only \cite{Verhoeven2015,Alboussiere2017,Jones2022}.  In this study, we take $\epsilon= 0.1$ and $D= 0.8$, the latter of which is still somewhat smaller than the maximum $D_{\rm max}=1-\epsilon$ at fixed $\epsilon$.
   \item {\em Fully compressible convection} exists for $\epsilon\approx D \approx 0.5$ and will be denoted to as FCC. It is the range in which the free-fall Mach number is maximum. In this study, we take $\epsilon= 0.45$ and $D= 0.5$. 
   \item {\em Strongly superadiabatic convection} exists for $D\ll 1$ and $\epsilon\to 1$ and will be abbreviated by SAC. In this study, we take $D= 0.1$ and $\epsilon= 0.8$, the latter of which is still somewhat smaller than the maximum $\epsilon_{\rm max}=1-D$ at fixed $D$.
\end{enumerate}
These four regimes of compressible convection are marked in the triangular $\epsilon-D$ phase plane in Fig. \ref{Fig1} (a). The temperature ratio, $ T_{\rm bot}/T_{\rm top} =1/( 1 - ( \epsilon + D))$  varies from 1.25 to 20 in this study as listed in table \ref{tab:simusupp} where further details on all simulations are given. For all simulations, $T_{\rm bot}$ is kept constant. For the physics considered  in this study,  the absolute value of $T_{\rm bot}$ is not relevant.
%-------------------------------------------------
%-------------------------------------------------

It should be noted that $M_{f} \rightarrow 0$  when either $\epsilon$ or $D$ go to zero. Our interest is in truly genuine compressible behavior limits  with non-negligible Mach number. Thus, we have chosen  finite values of $\epsilon=0.1$ and $D = 0.1$ for SSC and SAC cases, respectively. From  table \ref{tab:simusupp} it is seen that the obtained maximum turbulent Mach numbers are high enough such that   approximations including the pseudo-incompressible approximation \cite{durran1989,klein2012}, the low Mach number approximation of \cite{almgren2006},  or the anelastic \cite{ogura1962} approximation would not be valid. Indeed, finding out the ranges of validity of these approximations in  
the $\epsilon-D$ parameter plane and to study the transitions from these approximations to the fully compressible regime would be interesting and important. However, this is outside the scope of the present study; a brief discussion is included in the final discussion.

We furthermore note that the regimes SAC and FCC are relatively unexplored in the current literature. This is different to the SSC regime which has been studied frequently in the past, e.g., in refs. \cite{cattaneo1991,Verhoeven2015}. To the best of our knowledge, the present study extends the range of stratification in the SSC. In table \ref{tab:simusupp}, we provide the ratio of the  background density stratification which is given by 
%--------------------------------------------------
\begin{equation}
\frac{\bar\rho_{\rm bot}}{\bar\rho_{\rm top}}= (1 - D)^{-1/(\gamma-1)}\,,  
\end{equation}
%--------------------------------------------------
for all simulations. For the SSC case, this ratio is here about $56$  compared to about 20 and 11 in Verhoeven \textit{et al.} \cite{Verhoeven2015}  and Cattaneo \textit{et al.} \cite{cattaneo1991}, respectively. In ref. \cite{JPJSJFM2023}, we showed a transition to the highly stratified convection when the dissipation number exceeds $D_{\rm crit} \simeq 0.65$ (and $\bar\rho_{\rm bot}/\bar\rho_{\rm top}  \ge 13.8$) for fixed $\epsilon \approx 0.1$. For the largest dissipation numbers the pressure rather than the temperature fluctuations get synchronized with density fluctuations. Here in our SSC case, $D=0.8 > D_{\rm crit}$, thus our study will shed further light on the SSC limit case which includes a   detailed analysis of the differences between the top and bottom boundary layers.

\section{Set of Direct numerical simulations}

A total of 8 DNS were conducted at two different Rayleigh numbers of $Ra \approx 10^5$ and $10^6$ at $Pr=0.72$ for each of the four regimes with their operating points $(\epsilon,D)$, see again table \ref{tab:simusupp}. In addition, we add two corresponding Rayleigh-B\'{e}nard convection runs for reference, runs 1$_{\rm RBC}$ and 5$_{\rm RBC}$. The dimensionless parameters, $Ra$ and $Pr$ are given by 
\begin{equation}
Ra=\frac{\epsilon \bar\rho^{2}_{\rm bot} C_p g H^3}{\mu_0 k_0} \quad\mbox{and}\quad Pr=\frac{\mu_0 C_p}{k_0}.
\end{equation} 
 For all simulations, we use an aspect ratio $\Gamma= L/H = 4$, where $L$ is the horizontal length. Previous studies in isotropic compressible turbulence \cite{JDJFM2016}  have shown that a (uniform) grid resolution of $\eta/\Delta x \ge 0.5$ is sufficient to resolve the small-scale statistics correctly, such as moments of vorticity and dilatation up to fourth order. Here, $\eta$ is the Kolmogorov length scale. In our case, the wall-normal grid spacing, $\Delta z(z)$, and the Kolmogorov length scale, $\eta(z)$, are functions of depth $z$. The latter is given by 
\begin{equation}
   \eta( z ) =  \left[\frac{\mu_{0}^{3}}{\rho^3_{\rm ref}(z) \langle \epsilon_{f} (z) \rangle_{A,t}}   \right]^{1/4},
\end{equation}
where  $\rho_{\rm ref}(z)=\langle\rho \left(z\right) \rangle_{A,t}$.  The notation $\langle \cdot\rangle_{A,t} $  corresponds to a combined average over the horizontal directions and time which is given for a field $X$ as 
\begin{equation}
    \langle X (z) \rangle_{A,t} = \frac{ \sum^{N_{t}}_{m=1} 
    \sum^{N_{x}}_{i=1} \sum^{N_{y}}_{j=1} X(x_i, y_j,z,t_m)}{N_{x}N_{y}N_{t}},
\end{equation}
where $N_{x}$, $N_{y}$, and $N_{t}$ are the grid point numbers with respect to $x$-, $y$-directions, and number of snapshots, respectively.

\begin{figure}[t]
\centering
\includegraphics[width=0.4\linewidth]{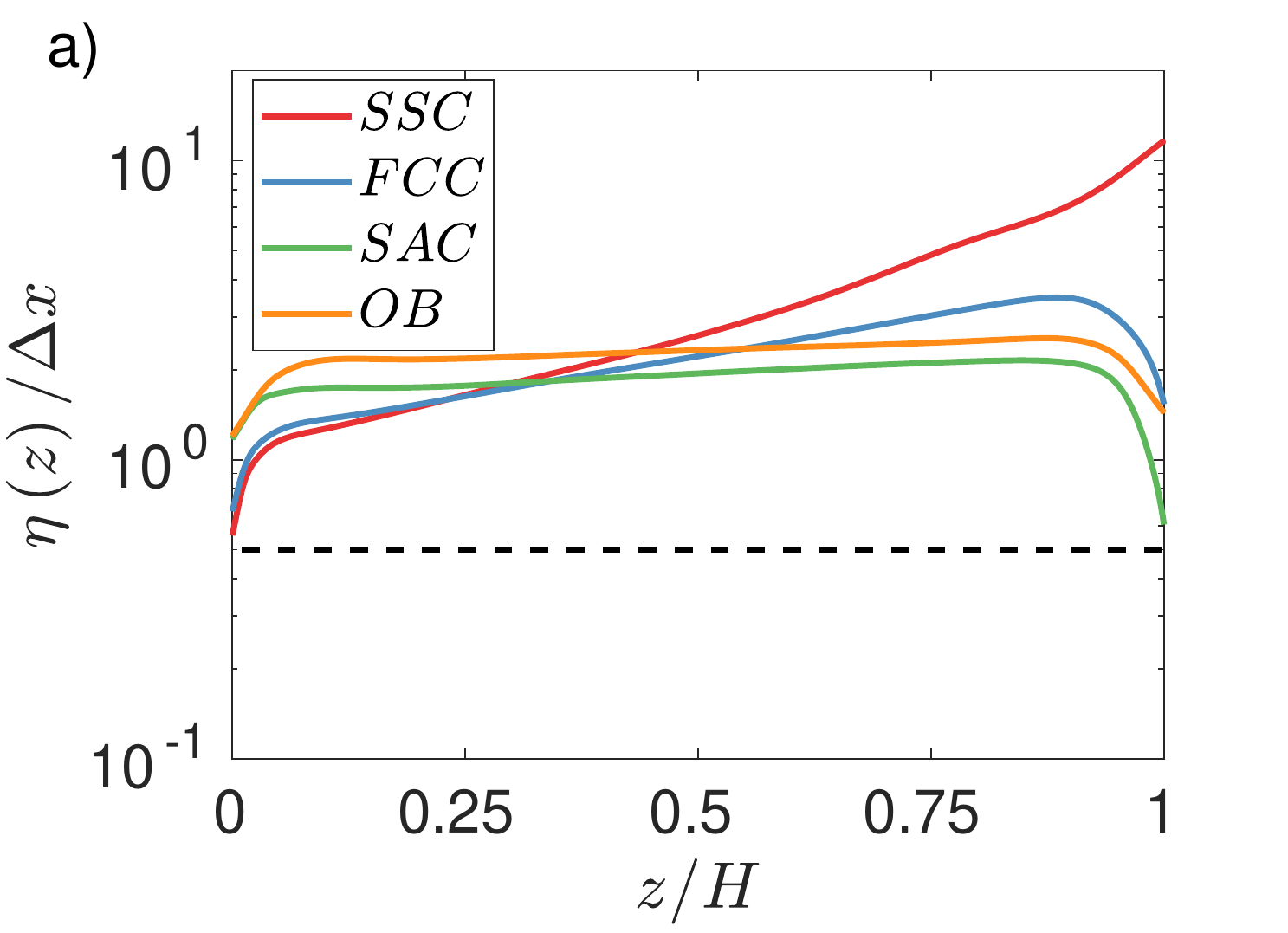}
\includegraphics[width=0.4\linewidth]{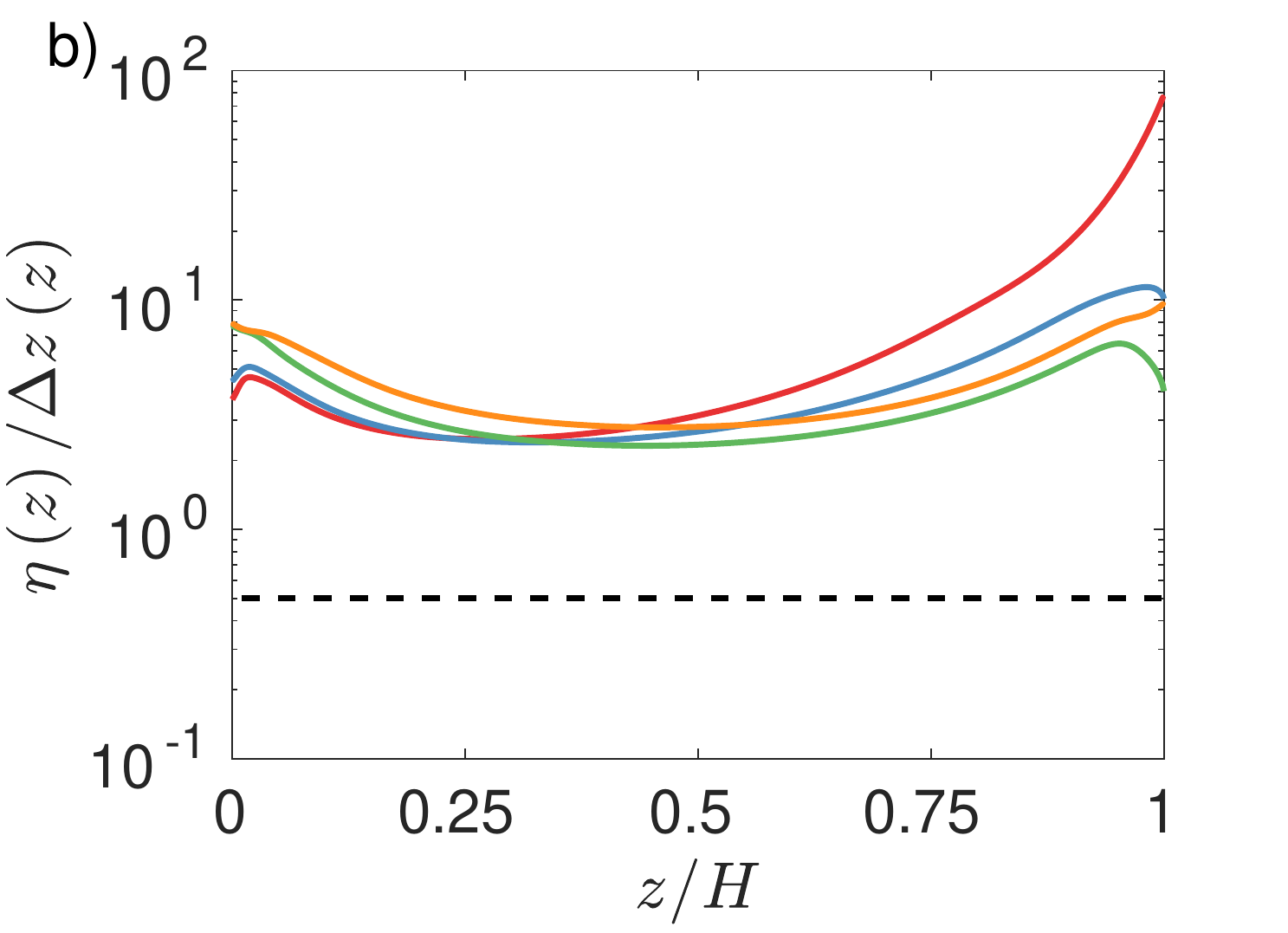}
\caption{Resolution of the DNS. Ratio of the height-dependent Kolmogorov length scale to the uniform horizontal grid size (a) and the non-uniform wall-normal grid size (b). In both figures, the black dashed line corresponds to the resolution criterion from ref. \cite{JDJFM2016} at a ratio of 0.5. Data are for $Ra=10^6$.}
\label{Fig0}
\end{figure}

The mean kinetic energy dissipation rate at height $z$ is given by 
\begin{equation}
\langle \epsilon_{f} (z)\rangle_{A,t} = \frac{\mu_{0}}{\langle \rho \rangle_{A,t}} \langle S_{ij} \sigma_{ij}\rangle_{A,t}\,.
\end{equation}
The horizontal and vertical resolution, $\eta/\Delta x$ and $\eta/\Delta z$, with respect to depth are shown for $Ra \approx 10^6$ in panels (a) and (b) of Fig. \ref{Fig0}, respectively. It is clear from both figures that for all cases, our  simulations are very well resolved, the minimum ratio is above the threshold. The grid resolution in the wall-normal direction (b) is better than the one in the horizontal  directions (a), due to much finer grid in $z$-direction. In table \ref{tab:simusupp}, we provide the number of grid points inside the top and bottom thermal boundary layers for all the simulations. The thermal boundary layer thicknesses at the bottom and top are therefore defined as
\begin{equation}
\lambda_{T}^{\rm bot,top}= \Big| \langle T_{\rm sa}(0,1) - T_{\rm sa}(0.5)\rangle_{A,t} \Big | \Bigg| \frac{\partial \langle T_{\rm sa} \rangle} {\partial z}\Bigg|^{-1}_{z/H=0,1}\,.
\label{locBL1}
\end{equation}     
Here $T_{sa}$ is the superadiabatic temperature which is given by
\begin{equation}
    T_{\rm sa} \left(\mathbf{x},t\right)  = T\left(\mathbf{x},t\right) - \overline{T} \left(z\right).
\end{equation}
We can conclude that both boundary layers are very well resolved which we also document in the last two columns of table \ref{tab:simusupp}, see also \cite{Scheel2013}. The present simulations are thus better resolved than those in \cite{Verhoeven2015}.

%-------------------------------------------------
%-------------------------------------------------

\section{Simulation Results}
\subsection{Turbulent heat and momentum transfer -- Nusselt and Reynolds number}
The Nusselt and Reynolds  numbers, $Nu$ and $Re$, are the global responses of the system to the parameters $Ra$, $Pr$, $D$, and $\epsilon$. They quantify the turbulent heat and momentum transfer. The Nusselt number is defined on the basis of the superadiabatic temperature field and is  given by
\begin{equation}
        Nu = \left . - \frac{H}{ \Delta \langle T_{sa} \rangle _{A,t}    } \frac{d \langle T_{sa} \rangle _{A,t}}{dz} \right|_{z=0,H} = \left . - \frac{H}{\epsilon T_{B}} \frac{d \langle T_{sa} \rangle _{A,t}}{dz} \right|_{z=0,H}. 
        \label{eq:nuss}
\end{equation}
The reported Nusselt number is the mean of those calculated at the top and bottom boundaries. The difference between both values was found to be negligible for all cases, with a range between $0.06 \%$ and  $3\%$. This indicates that our simulations are well-converged and statistically steady. The Reynolds number is defined as 
\begin{equation}
    Re = \frac{\rho_{\rm ref} u^{\prime} H }{\mu_0} \quad \mbox{with} \quad u^{\prime}=\sqrt{\langle u_x^2\rangle_{V,t}+\langle u_y^2\rangle_{V,t}+\langle u_z^2\rangle_{V,t}}\,.
\end{equation}
The Nusselt and Reynolds numbers for all cases  are listed in table \ref {tab:simusupp}. For  a given $Ra$, the SAC case has the highest Nusselt and Reynolds numbers whereas the SSC case the lowest. The values of the FCC cases are in between those of SAC and SSC.  In fact, they are close to the OB case. In general,  heat and momentum transfer efficiencies of the fluid system are enhanced and decreased by increasing $\epsilon$ and $D$, respectively. As expected, the Nusselt  number of both OB-like cases is close to that of the corresponding incompressible Rayleigh-B\'{e}nard case. The slight difference between the two can be accounted for the fact that small, but non-negligible compressibility effects are expected even when $\epsilon= D= 0.1$.

\subsection{Structure of the superadiabatic temperature field}
Although the mean values of Nusselt and Reynolds numbers of FCC and OB are comparable, we observe significant differences in contour plots of the superadiabatic temperature for all cases including FCC and OB in  Figs \ref{Fig1}(b)--(e). Panel (b) displays the OB-like dynamics with symmetric plume detachments from the boundary layers at top and bottom; the simulations of SAC in (c) develop a top-down asymmetry between falling and rising plumes which will still not significantly alter the mean profiles across the layer as we will see further below. Highly asymmetric configurations follow for FCC at the operating point $(\epsilon,D)$ close to the Mach number maximum $M_f^{\ast}$ in panel (d) and for SSC with $D\to 1-\epsilon$ where turbulence is strongly suppressed at the top and a clearly visible sublayer with reduced turbulence is formed, see Fig. \ref{Fig1}(e). Except for the OB case, thinner plumes emanate from the top compared to those rising from the bottom. This is a  general trait of compressibility regardless of the  boundary layer thickness. Moreover, comparing the large $D$ (FCC and SSC) cases, the stratified layer at the top becomes prominent for SSC at $D= 0.8$ compared to FCC at $D= 0.5$. Also, the plumes which detach from the bottom in FCC get well-mixed compared to those in SSC due to higher $\epsilon$.

%----------------------------------------------------------------
\begin{figure}[t]
\centering
\includegraphics[width=0.32\linewidth]{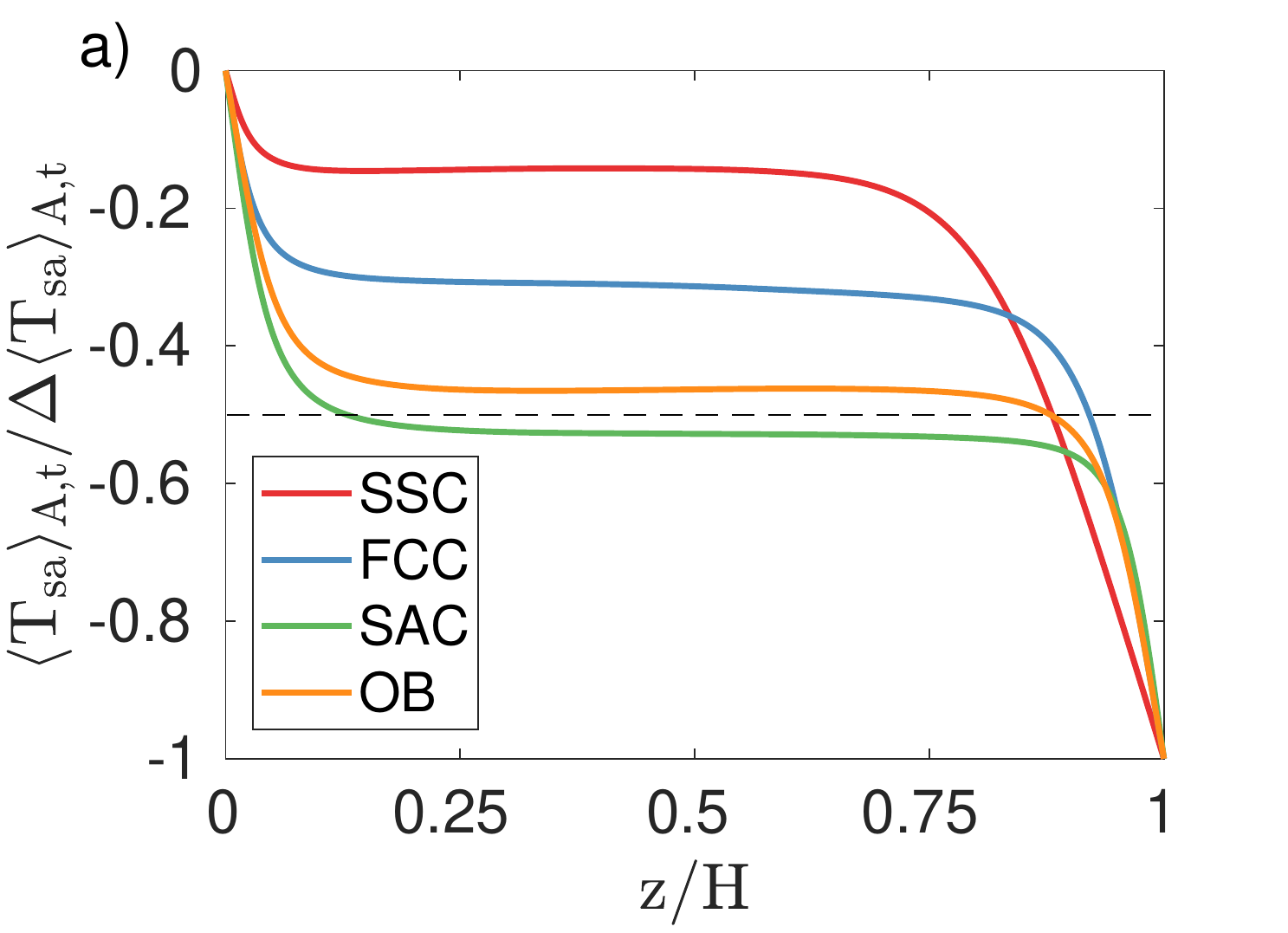}
\includegraphics[width=0.32\linewidth]{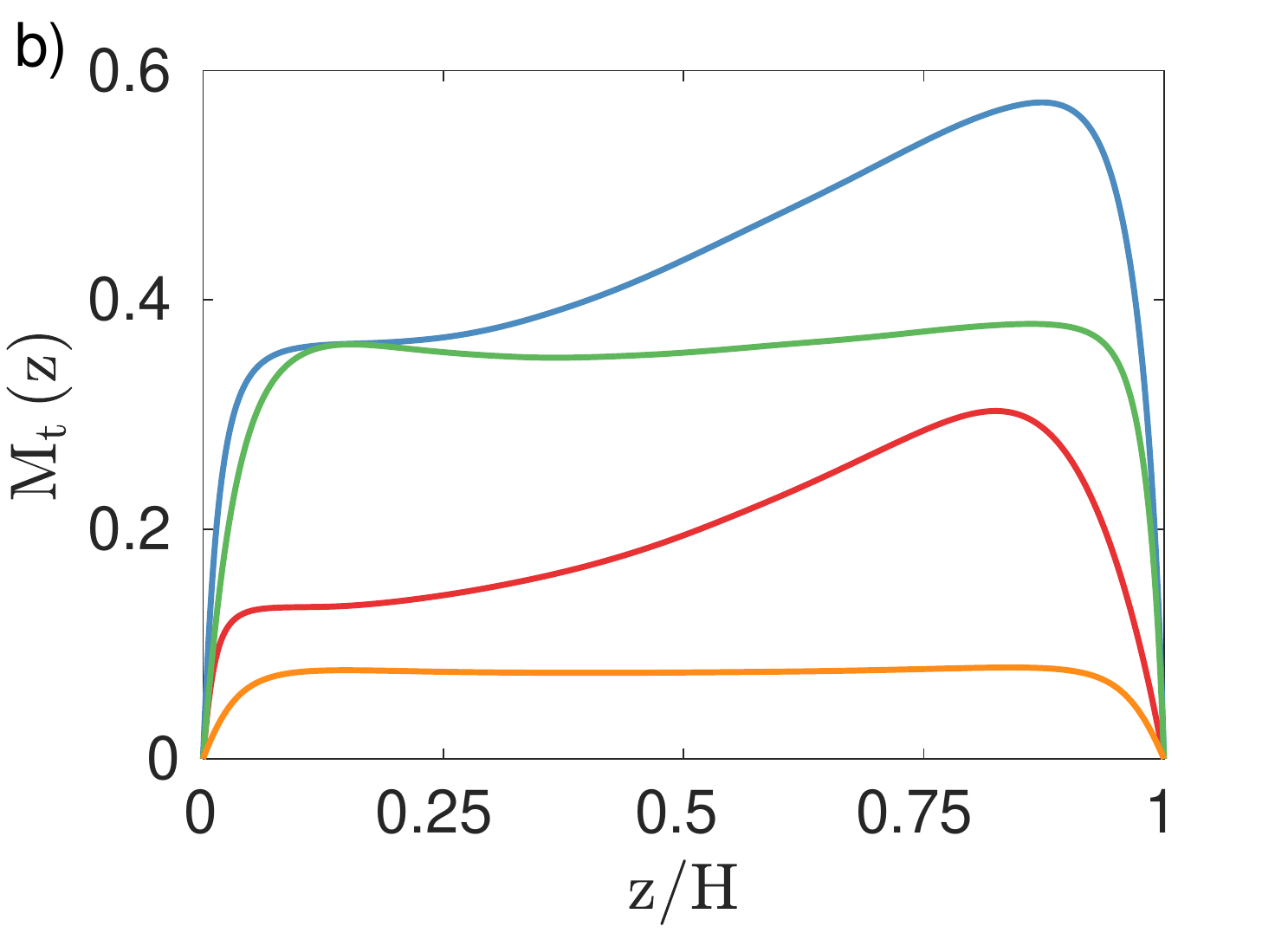}
\includegraphics[width=0.32\linewidth]{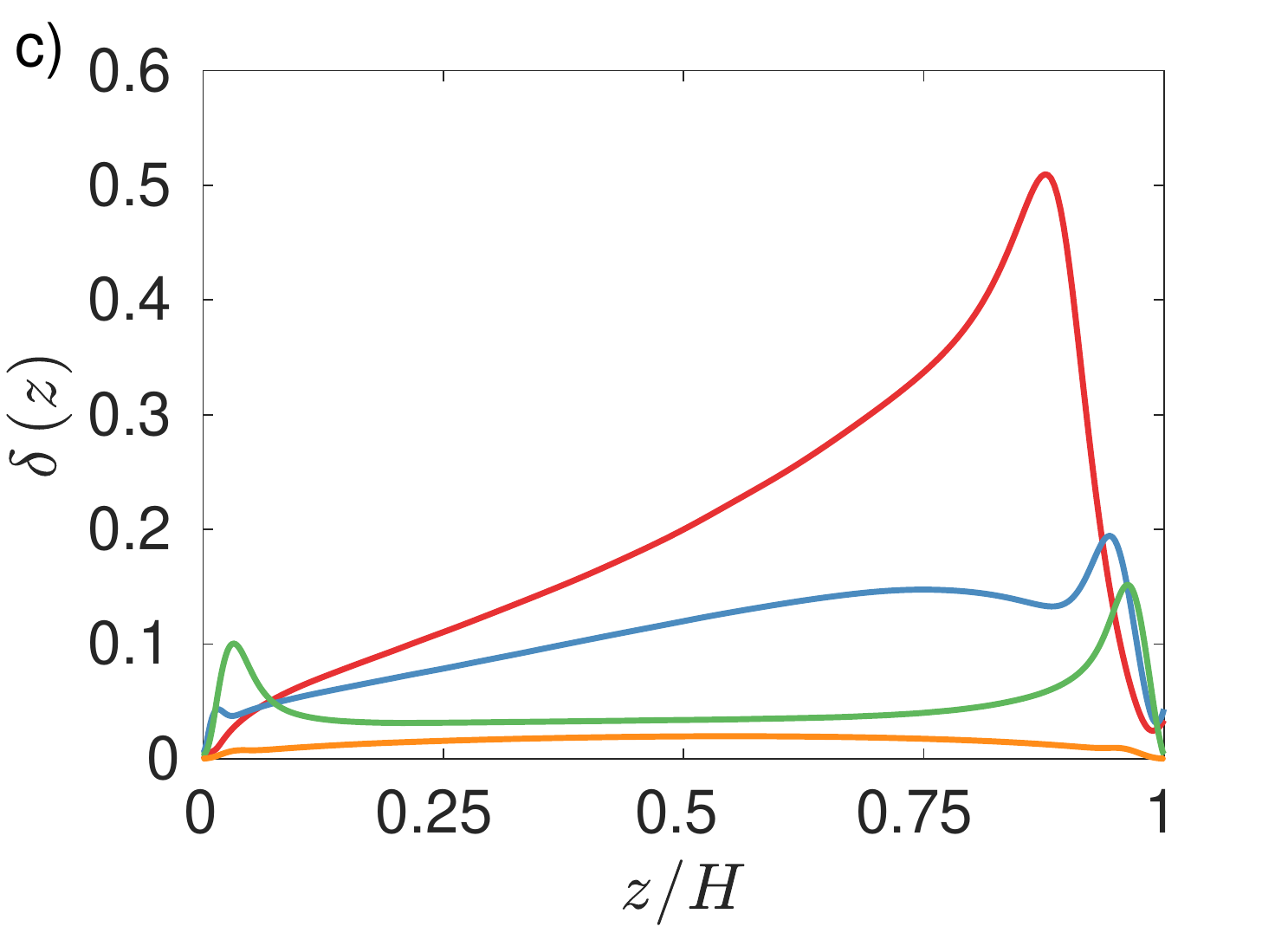}
\includegraphics[width=0.32\linewidth]{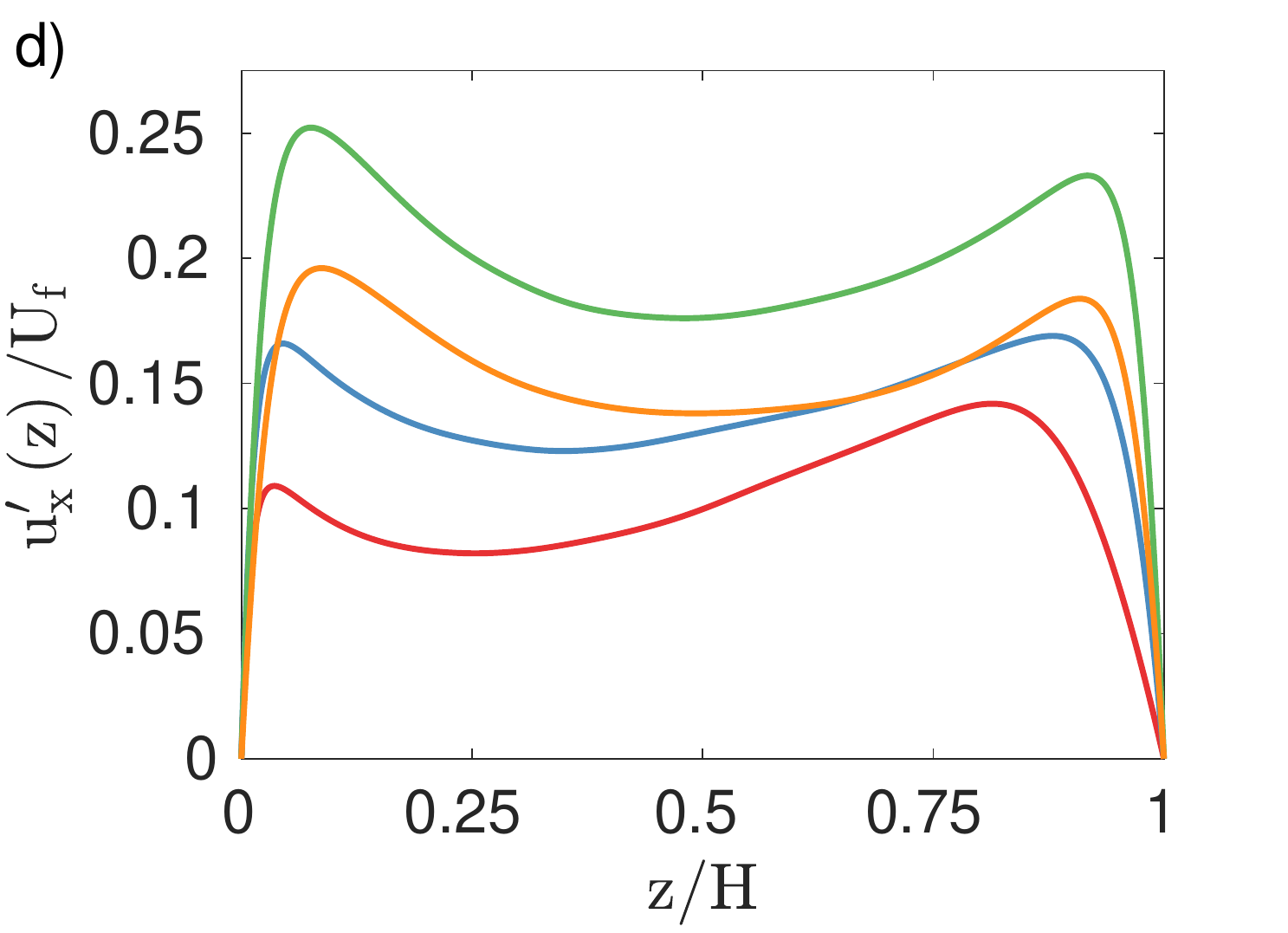}
\includegraphics[width=0.32\linewidth]{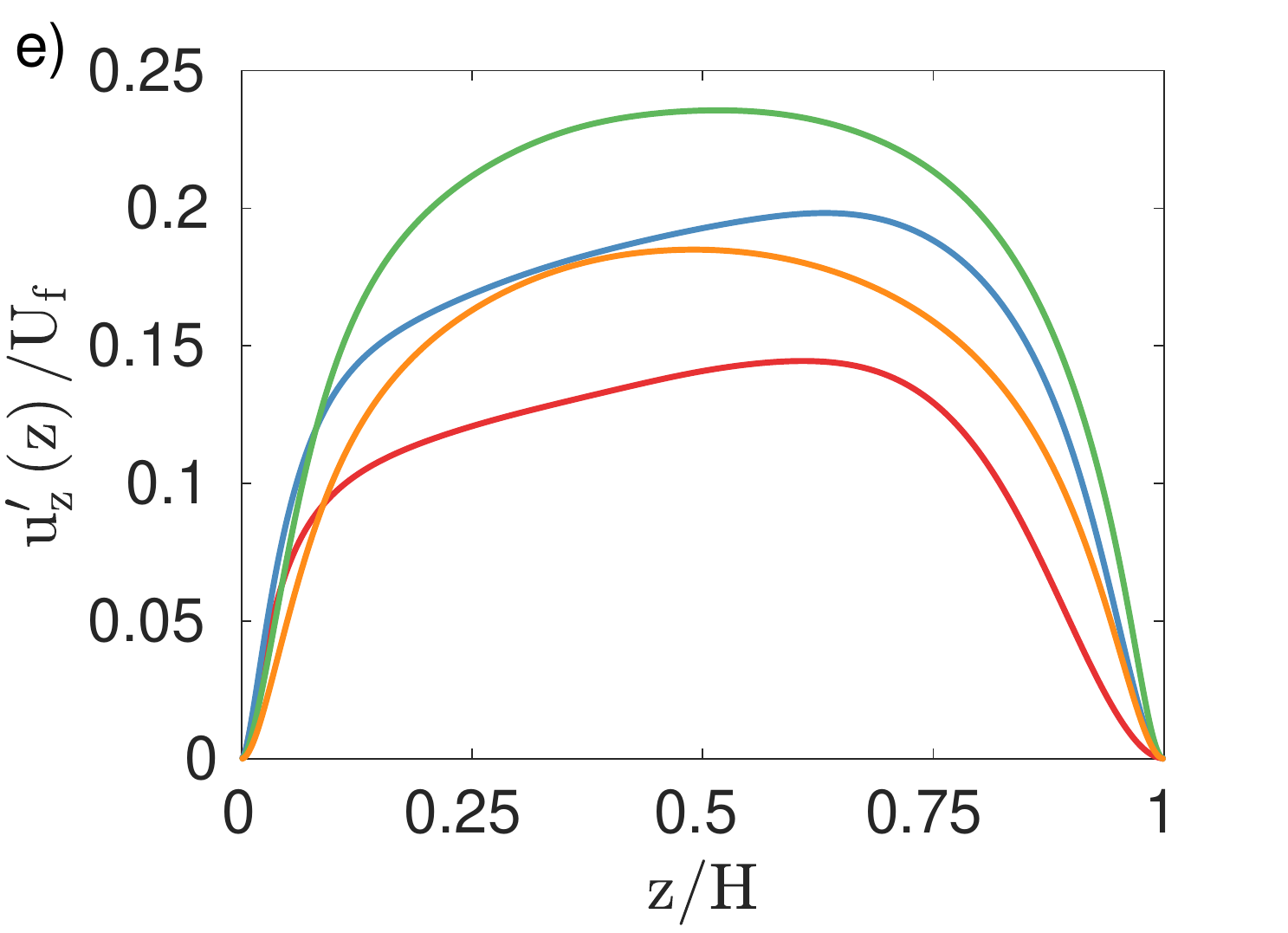}
\includegraphics[width=0.32\linewidth]{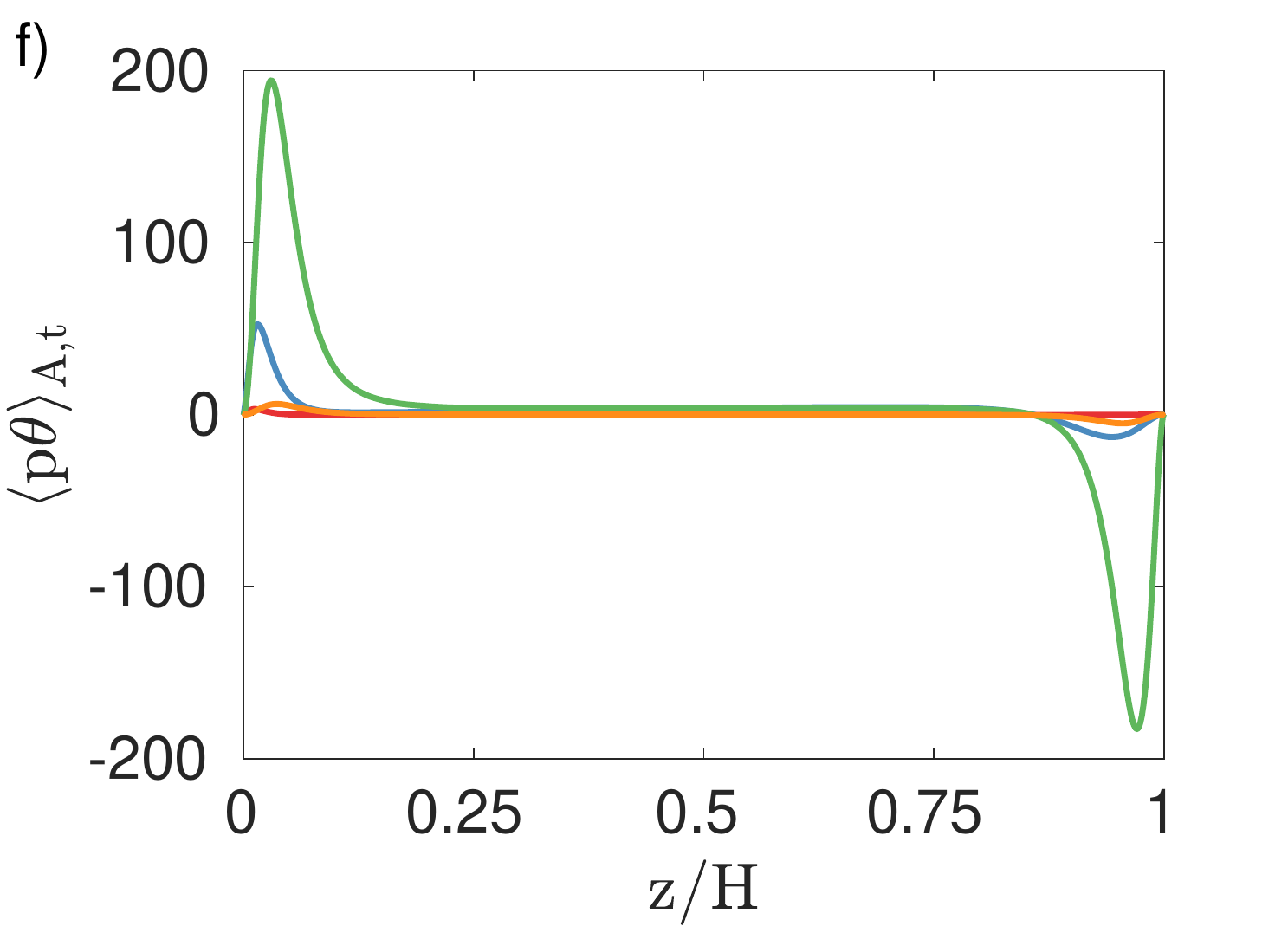}
\caption{Mean vertical profiles of plane-time averaged quantities as a function of superadiabaticity and dissipation number at $Ra\simeq 10^6$ (runs 5--8). We show  (a) the normalized mean of the superadiabatic temperature $T_{\rm sa}(z)$, (b) the turbulent Mach number $M_t(z)$, (c) the dilatation parameter $\delta(z)$, (d) the root mean square profiles of $u_x^{\prime}(z)$ and (e) $u_z^{\prime}(z)$, and finally in panel (f) the mean of the pressure dilatation, $p\theta=p({\bm\nabla}\cdot{\bm u})$. The color coding for all plots corresponds to the legend in panel (a). Data are for $Ra=10^6$.}
\label{Fig2}
\end{figure}
%----------------------------------------------------------------

\subsection{Vertical mean profiles and breakdown of top-down symmetry}
The mean vertical profiles of the turbulence fields are summarized in Fig.~\ref{Fig2}. We recognize first that for most statistics, the FCC profiles are enclosed by those of the limit cases, SAC and SSC, which supports the idea of being a blend of these limits. Panel (a) shows the combined plane-time averages of the superadiabatic temperature $T_{\rm sa}$ renormalized by $\Delta_{\rm sa}=\langle T_{\rm sa}(0)\rangle_{A,t}-\langle T_{\rm sa}(1)\rangle_{A,t}$ such that all four profiles are comparable. Cases OB-like and SAC with $D\ll 1$ give nearly symmetric profiles with a well-mixed bulk close to $\langle T_{\rm sa}\rangle / \Delta_{sa} = 0.5$. In contrast, the cases FCC and SSC are characterized by significant offsets from the symmetric OB value of $\langle T_{\rm sa}(z)\rangle_{A,t}=0.5$. However, since $\epsilon$ and $D$ are finite and not zero in the OB-like case (and thus the compressibility effects are present) we observe a small positive offset from a perfect top-down symmetry. In SAC, for which the superadiabaticity gets very large while $D$ remains unchanged, the cold falling thinner plumes reach deeper into the bulk. This reverses the slight offset of the OB-like  case into a negative one, $\langle T_{\rm sa}(z)\rangle_{A,t} < 0.5$.

Mean vertical profiles of the {\em turbulent Mach number} $M_t$, defined as 
\begin{equation}
M_t(z)=\frac{u^{\prime}(z)}{\sqrt{\gamma R \langle T \left(z \right) \rangle_{A,t}   }}
\end{equation}
 are reported in Fig.~\ref{Fig2}(b). We observe again  asymmetric curves for FCC and SSC in agreement with those for the velocity fluctuations that will be discussed afterwards. The largest turbulent Mach numbers are found in FCC close to the top with values that remain consistently below estimate, $\max_z (M_t(z)) < M_f^{\ast}$ from \eqref{def2}. The smallest values of $M_t\left(z\right)$ are observed for the OB-like regime. The estimate $M_f\le 1$ in \eqref{def2} was based on the bottom temperature $T_{\rm bot}$. A free-fall Mach number based on top plate temperature could alternatively be defined, such that two free-fall Mach numbers are obtained as a reference,
\begin{equation}
  M^{\rm bot}_f=M_f\quad \mbox{and}\quad M^{\rm top}_{f}=  \sqrt{  \frac{\epsilon D}{\left(\gamma - 1 \right)}}  \sqrt{  \frac{T_{\rm bot}}{T_{\rm top}}}\,.  
\label{newmach}
\end{equation}
For the FCC case, the free-fall Mach number with respect to the top is $M^{\rm top}_{f}= 3.35$; thus in principle supersonic conditions can exist at the top boundary, see also \cite{Cattaneo1990}. In  table \ref{tab:simusupp}, we list the maximum (pointwise) turbulent Mach number $M_t^{\rm max}=\max_{({\bm x},t)} |{\bm u}({\bm x},t)|/[\gamma R T({\bm x},t)]^{1/2}$ obtained in our simulations for all cases. As expected, the highest maximum Mach number is for the FCC case around $1.32$ for $Ra \approx 10^{6}$. Although supersonic, this is much lower than $M^{\rm top}_{f}= 3.35$. This apparent discrepancy is because, near the top boundary, the velocities are negligible and the strongest velocity fluctuations are near the edge of the top  boundary layer where the temperature in turn is higher than $T_{\rm top}$.   Indeed, we also see from table \ref{tab:simusupp}, for all cases, $M^{max}_{t}$ increases with the Rayleigh number (even though two $Ra$ values are available only).  Additional studies at higher Rayleigh numbers are thus required to understand the trend of compressibility effects with Rayleigh number.

In compressible turbulence, $M_{t}$ alone cannot characterize the system. Donzis and Panickacheril John \cite{Donzis2020} showed that by including a further parameter -- the {\em dilatation parameter} $\delta$ -- along with $M_{t}$, universality for essential compressible statistics \cite{Panickacheril2019, Panickacheril2021} was shown, thus an easier systematic analysis of a compressible turbulent system is possible. Similar to $Nu$ and $Re$, the parameters $M_t$ and $\delta$ are the responses that show deviations from the Boussinesq regime, which are set by superadiabaticity $\epsilon$ and dissipation number $D$. In detail, the new parameter $\delta$ is given by 
\begin{equation}
\delta=\frac{u_d^{\prime}}{u_s^{\prime}} \quad \mbox{with}\quad {\bf u}={\bf u}_s+{\bf u}_d\,,
\end{equation}
where the velocity field is subject to a Helmholtz decomposition into solenoidal and dilatational parts. Such decomposition is less straightforward in an inhomogeneous flow and works via the corresponding dissipation rates. We therefore assume the scaling of $\langle \epsilon_{d} \rangle/ \langle \epsilon_{s} \rangle \propto \delta^{2}$, where $\langle \epsilon_{d} \rangle  = 4\nu_0/3 \langle (\mathbf{\nabla}  \cdot \mathbf{u})^2 \rangle$  and $\langle \epsilon_{s} \rangle  = \nu_0 \langle\omega_{i}^2 \rangle$ are the mean dilatational and solenoidal kinetic energy dissipation rates \cite{Sarkar1989}, respectively. Here, $\omega_{i}$ is the vorticity vector field and $\nu_0=\mu_0/\rho_{\rm ref}$ the kinematic viscosity. For homogeneous flow, $\langle \epsilon_{f} \rangle  =  \langle \epsilon_{s} \rangle  + \langle \epsilon_{d} \rangle$, but in our case, there will be inhomogeneous contributions  to $\langle \epsilon_{f} \rangle$. However, even for our cases, the ratio of $\langle \epsilon_{d} \rangle/ \langle \epsilon_{s} \rangle $  will give us a relative estimate of the dilatational motions between the different cases, we are considering here.  Thus, we define 
\begin{equation}
    \delta (z)  = \sqrt{ \frac{\langle\epsilon_{d}(z) \rangle_{A,t}}{\langle\epsilon_{s} (z) \rangle_{A,t}} } = 2 \sqrt{\frac{\langle (\mathbf{\nabla}  \cdot \mathbf{u})^2 \rangle_{A,t}}{3\langle \omega_{i}^2 \rangle_{A,t}}} . 
    \label{eq:deltadef}
\end{equation}
In our context, $\delta(z)$ can be interpreted as the relative strength of kinetic energy dissipation due to shocklets or pre-shocks to that due to the vortical fluid motions. The higher $\delta$, the larger fractions of acoustic motions in the flow field.

We show the variation of $\delta(z)$ in Fig. \ref{Fig2}(c). Once again, the asymmetry between the top and bottom increases for growing $D$, with SSC being the case with dominant dilatational (or compressible) motion at the top boundary followed by the FCC case. For smaller $D$ (in SAC and OB), the magnitude of the dilatational motions is more or less comparable at the top and bottom. Comparing the cases OB and SAC, the effect of superadiabaticity is to increase the strength of dilatational motions at both top and bottom in equal proportions. For high dissipation number $D$, i.e., the cases FCC and SSC, the parameter $\delta$ decreases monotonically from the top to the bottom in the bulk region of the convection layer.

The normalized mean profiles of the root mean squares of two velocity components are shown in Figs. \ref{Fig2}(d,e). The profile of the horizontal component $u_x^{\prime}(z)$ in Fig. \ref{Fig2}(d) are qualitatively similar for all cases. The magnitude decreases with $D$, but increases with $\epsilon$. The profiles of the $y$-component is similar and thus not shown.  The vertical component $u'_{z}(z)$ is shown in Fig. \ref{Fig2}(e). While the mean profiles are nearly symmetric for OB and SAC again, fluctuations are enhanced towards the top in the other two cases, FCC and SSC.  

For compressible turbulence,  the entropy is the conserved quantity rather  than $T_{sa}$.  Apart from temperature, there is an additional contribution to  entropy from the pressure dilatation correlation, $p\theta$, where  $\theta({\bm x})={\bm \nabla}\cdot {\bm u}$ is the divergence of the velocity field. We plot $\langle p\theta \rangle_{A,t}$ in Fig. \ref{Fig2}(f). One finds that this term is negligible for the OB and SSC cases,  where $\epsilon \ll 1$. The pressure dilatation increases with $\epsilon$ as seen for the FCC and SAC cases. However, significant variations are observed near the boundaries only and remain negligible in the bulk. This bulk behaviour is in fact  consistent with homogeneous isotropic compressible turbulence \cite{sarkar1992}. Although the magnitude of the instantaneous pressure dilatation fluctuations are considerably high, due to its oscillatory nature, the  net contribution to the energy balance  after  averaging is negligibly small. The bulk behavior in convection is to a certain extent comparable to isotropic turbulence.   Jones \textit{et al.} \cite{Jones2022} estimate a positive offset of $T_{sa}$ in the bulk. However, empirical evidence from this study suggests that the offset of $T_{sa}$ in the bulk remains small as observed in Figs. \ref{Fig2}(a) and (f).

\subsection{Velocity field divergence} 

\begin{figure}[t]
\centering
\includegraphics[width=0.7\linewidth]{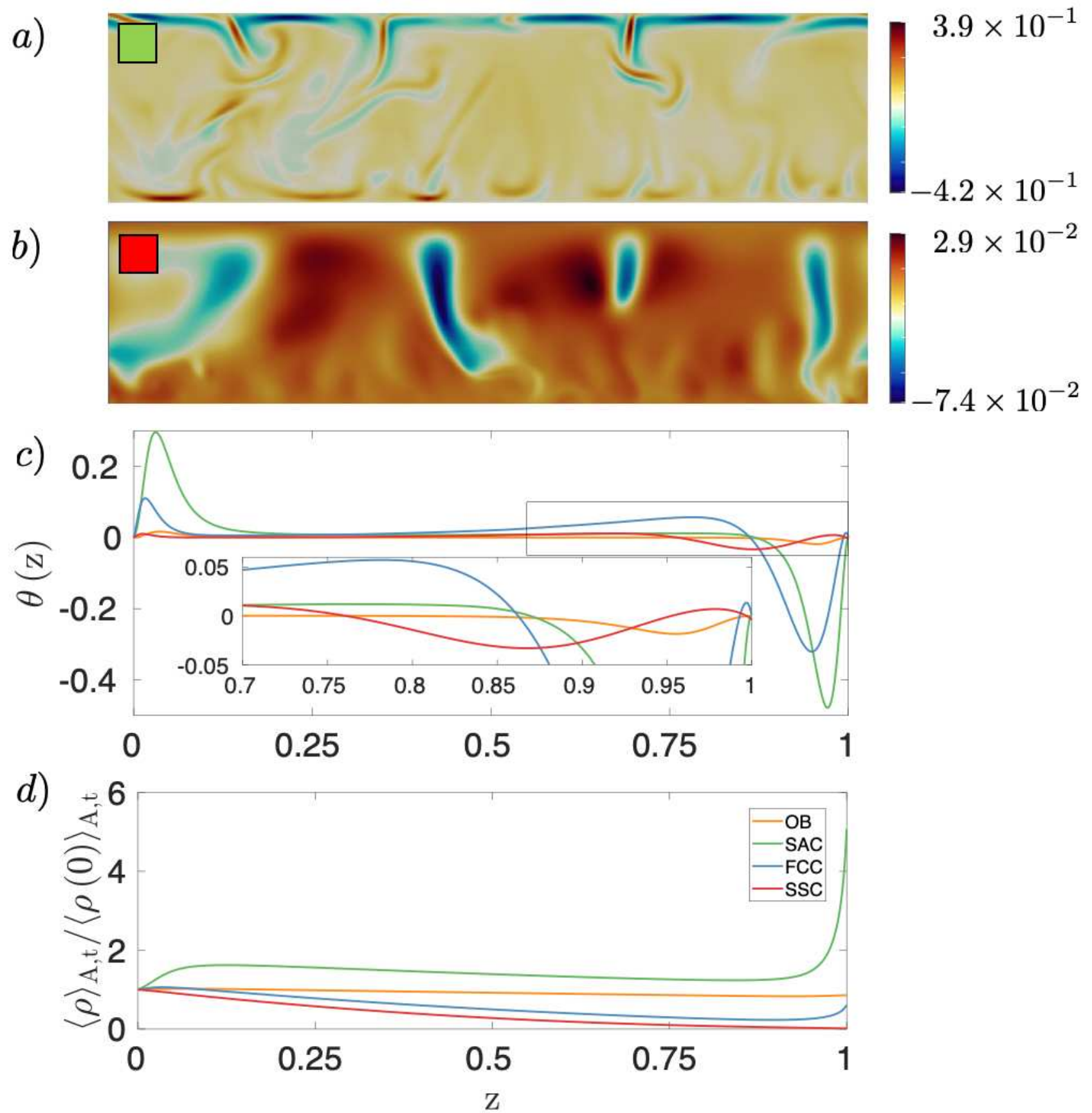}
\caption{Velocity divergence and mass density analysis. Instantaneous vertical cut of ${\bm \nabla}\cdot {\bm u}$ for (a) strongly superadiabatic case (SAC), which corresponds to the snapshot in Fig. \ref{Fig1}(c), and (b) strongly stratified case (SSC), which corresponds to Fig. \ref{Fig1}(f). (c) Mean vertical profiles of $\theta(z)=\langle {\bm \nabla}\cdot {\bm u} \rangle_{A,t}$. The inset magnifies the profiles close to the top wall as indicated by the box in the main panel. (d) Mean vertical mass density profiles. Again $Ra\simeq 10^6$ and $Pr=0.7$ in all cases. Line styles in panels (c,d) are identical to Fig. \ref{Fig2}.}
\label{Fig3}
\end{figure}

Figure \ref{Fig3}(a) shows an instantaneous cut through $\theta({\bm x})={\bm \nabla}\cdot {\bm u}$ for SAC which corresponds to Fig. \ref{Fig1}(c). Large regions of expansion ($\theta>0$) are found near the bottom boundary  which corresponds to  rising plumes.  The expansive regions near the top boundary also correspond to falling plumes, however they are surrounded by large regions of compression ($\theta<0$) from the upward flow. The compression effect at the top is more pronounced since $\langle\rho(z)\rangle_{A,t}$ grows strongly for $z/H>0.9$, see Figs. \ref{Fig3}(c,d) where we plot mean vertical profiles of the divergence $\theta(z)=\langle{\bm \nabla}\cdot {\bm u}\rangle_{A,t}$ and the density $\langle\rho(z)\rangle_{A,t}$. Maxima and minima of $\theta(z)$ are present in the bottom and top  boundary layers, respectively. Moreover, since the directions $x$ and $y$ are homogeneous,  $\theta(z)=\langle{\bm \nabla}\cdot {\bm u}\rangle_{A,t} = \langle \partial u_z/\partial z\rangle_{A,t}$. Note also that due to conservation of mass in our setup,  $\int_0^1\theta(z) dz=0$; therefore stronger compression at the top relative to expansion at the bottom results in a thinner top boundary layer.

The SAC limit is consequently weakly top-down-asymmetric only. The asymmetry  of the SAC case towards the top boundary, see Fig. \ref{Fig2}(a), can be partially explained as follows. Along with constant dynamic viscosity $\mu_0$ and thermal conductivity $k_0$, the high and low density at the top and bottom results in low and high kinematic viscosity $\nu=\mu_0/\rho$ and thermal diffusivity $\kappa=k_0/(c_p\rho)$, respectively. Similar to incompressible convection, this can lead to a thinner boundary layer at the top for the fluid with a higher density. However in SAC, strong compressible mechanisms lead to these strong density variations. 

We conclude that the stronger asymmetry of boundary layers in the FCC case (see Fig. \ref{Fig2}) -- the blend of SAC and SSC -- have to be traced back mainly to the SSC limit which is further detailed in Fig. \ref{Fig3}(b). In this regime the mass balance can be simplified to the anelastic case for sufficiently small $\epsilon$, ${\bm \nabla}\cdot(\bar\rho {\bm u})\approx 0$ \cite{Verhoeven2015}. (Strictly speaking, this relation is only valid for $M_{t} \ll 1$ which is not the case for the present SSC case  near the top boundary. The specific implications of the high compressibility  for $0.9< z/H <1$ were analysed in \cite{JPJSJFM2023} and will be briefly discussed in the final section. For most of the convection domain, anelasticity can be assumed in the SSC case.)  With a monotonically decreasing adiabatic profile $\bar{\rho}(z)$ this leads to a pointwise relation of
%------------------------------------------------------------------------
\begin{equation}
{\bm \nabla}\cdot{\bm u} = {\bm \nabla}_h\cdot{\bm u}_h+\frac{\partial u_z}{\partial z} \approx \Big|\frac{d\bar{\rho}}{dz}\Big|\frac{u_z}{\bar{\rho}}\,.
\label{AE}
\end{equation}          
%------------------------------------------------------------------------
\begin{figure}[t]
\centering
\includegraphics[width=0.47\linewidth]{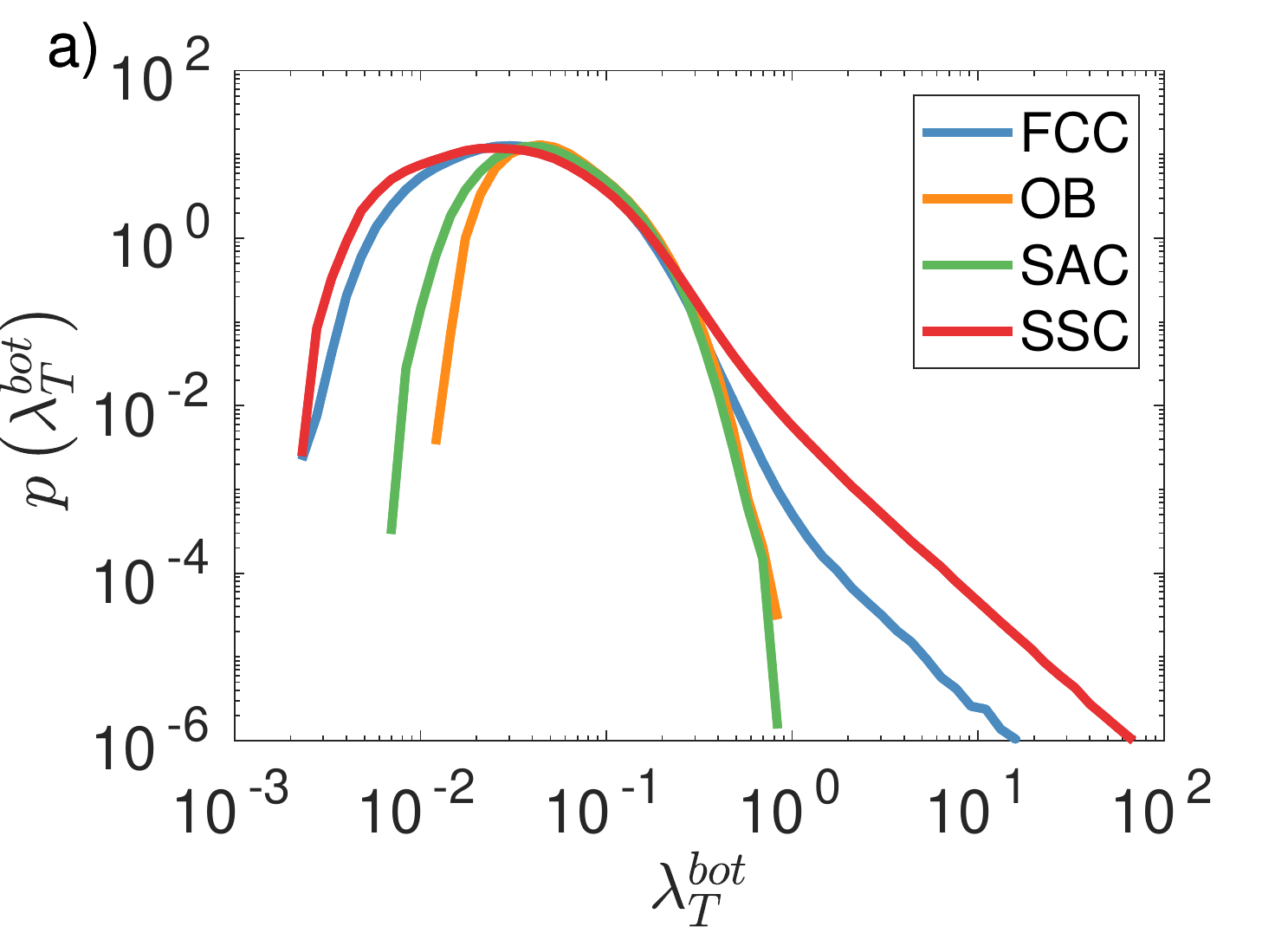}
\includegraphics[width=0.5\linewidth]{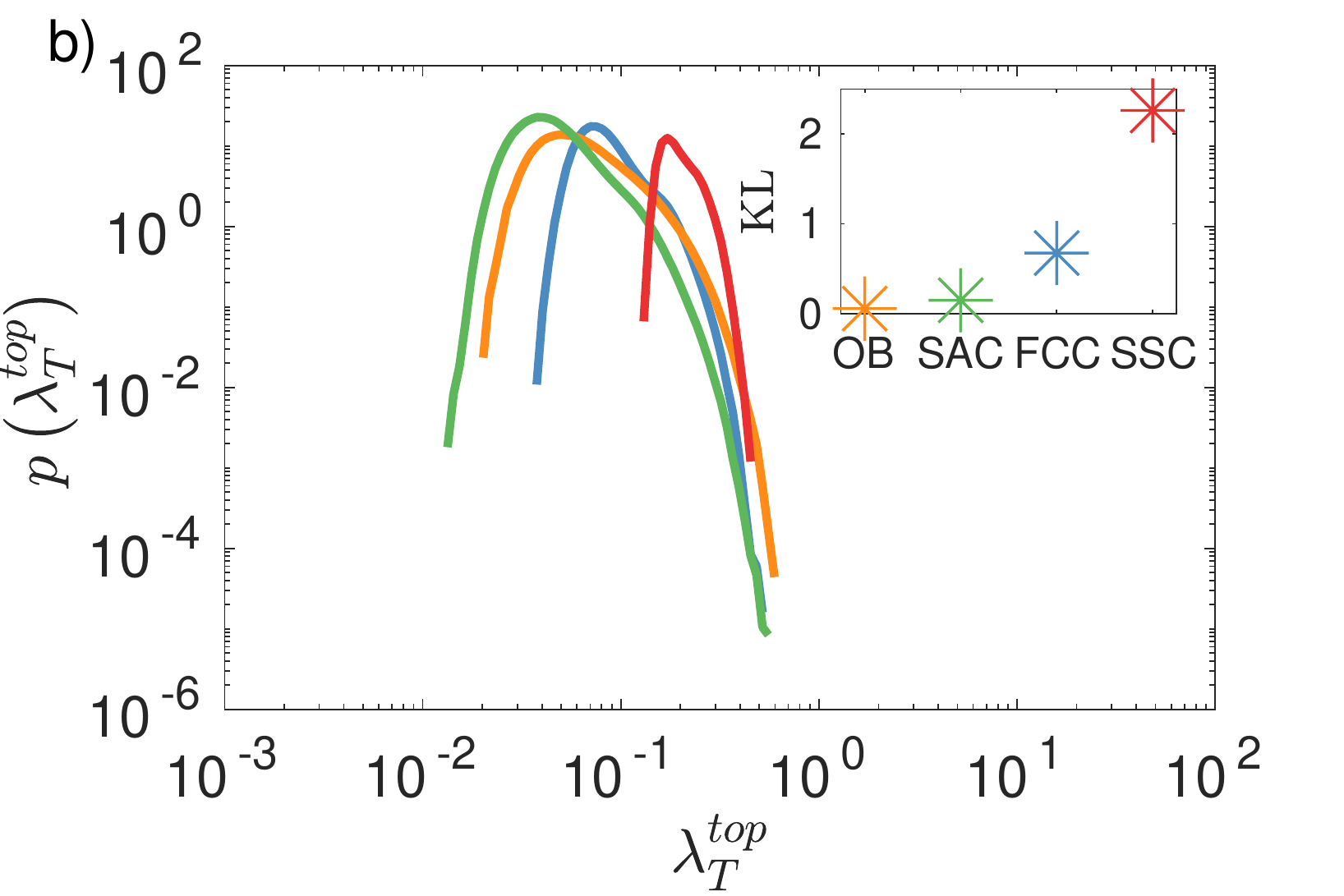}
\caption{Probability density functions of the local thermal boundary layer thickness at the bottom (a) and top (b). The inset in (b) displays the Kullback-Leibler divergence (KL) for the 4 runs at $Ra\approx 10^6$.}
\label{Fig4}
\end{figure}
%------------------------------------------------------------------------
Generally, any up-- and downwelling plume should expand and compress, respectively.  Depending on the sign of $\partial u_{z} / \partial z$, one gets structures which are different from the incompressible case. Close to the bottom, SSC has $u_z>0$ in plume detachment zones and thus $\theta({\bm x})>0$ by eq. \eqref{AE} since $\rho>0$. The horizontal velocity ${\bm u}_h$, that converges to these detachment regions, is weaker and thus less strictly converging as in the incompressible case. This leads to broader regions with accelerated upward flow, $\partial u_z/\partial z >0$, compared to incompressible convection.  It is clear from Fig. \ref{Fig3}(b) that the bottom boundary is dominated by expansions with $\theta ( \bm{x}) >0$ and $u_{z} > 0$ which results in a positive $\theta(z)$ near the bottom. The figure also shows that a positive velocity field divergence fills nearly the whole bulk, interrupted only by negative one with a significantly higher magnitude in the narrow downward plume regions. It can be expected that this contrast (which leads to a net zero mean as seen in the mean profiles) increases for $D\to 1-\epsilon$, see also a discussion in Spruit \cite{Spruit1997} in the solar context. This is not clear in Fig. \ref{Fig3}(c), since the  magnitude of $\theta \left( \bm{x} \right)$ for SAC and FCC with their larger $\epsilon$ is greater than in  SSC. Qualitatively, the bottom boundary dynamics of SAC and SSC is similar. 

At the top, following from Fig. \ref{Fig3}(c) (see also the inset), we get $\theta(z)<0$ for  $0.6\lesssim z/H \lesssim 0.9$, except directly near the wall for $z/H \gtrsim 0.9$. $\theta(z)<0$ implies upward  decelerating plumes coming from the bottom,  similar to  SAC. This broadly upwelling fluid is increasingly compressed which leads to $d\langle \rho_{\rm sa}\rangle_{A,t}/dz >0$ for $0.6\lesssim z\lesssim 0.9$ with $\rho_{\rm sa}=\rho-\bar{\rho}(z)$ (not shown). However, for SSC with its strong stratification, the plumes cannot reach the top boundary anymore. From Figs. \ref{Fig3}(b,c), it is clear that the magnitudes of $\theta (z)$ and $\theta( \bm{x})$ in the layer $0.9 \lesssim z/H \lesssim 1.0$ are smaller than for $0.6\lesssim z/H \lesssim 0.9$. The near-wall layer corresponds to very weak fluid motion due to high stratification. Here, thermal plumes detach from the top wall and get increasingly focused when shooting downwards deep into the bulk in agreement with \eqref{AE}. As seen from  Figs. \ref{Fig3}(b) and \ref{Fig1}(f), these downward-directed plumes always correspond to a negative divergence. This is in contrast to SAC, where the downwelling thermal plumes effectively expand even though they are compressed in the horizontal directions. 

Comparing Figs. \ref{Fig1}(d) with \ref{Fig3}(a) and Figs. \ref{Fig1}(f) with \ref{Fig3}(b) for SAC and SSC, respectively, we observe a one-to-one correspondence between thermal plumes from the top in $T_{sa}$ and high-magnitude regions of $\theta$. This suggests  genuine compressibility effects on the top boundary. 

For the SSC case in Fig. \ref{Fig3}(d), the density near the bottom is large compared to the top, which would imply a higher Rayleigh number near the bottom. However, we observe thinner plumes from top boundary due to compressibility. Thus high compressibility plays a significant role in plume   formation and structure.   The background  stratification of density  is observed in the bulk region for all four cases, see table \ref{tab:simusupp} . This is of course most obvious for the   cases with high $D$, the SSC $(D= 0.8)$ and FCC $(D= 0.5)$ cases.  The superadiabaticity $\epsilon$ is measure of departure from the equilibrium adiabatic profile. Thus for the OB and SSC cases, where $\epsilon= 0.1$, the departure from the adiabatic profile is  not apparent. A significant departure from the adiabatic profile is seen for the SAC case at both the boundaries  with $\epsilon= 0.8$ and $D= 0.1$.  Positive and negative density gradients  towards the wall are observed at the top and bottom boundaries respectively . They correspond to unstable convective configurations. Thus as expected, for the SAC case the high superadiabaticity of $\epsilon= 0.8$ easily overcomes  the weak  background stratification which corresponds to $D= 0.1$.  Consistent with thinner plume structures seen in Figs. \ref{Fig1} (c)  and  \ref{Fig3} (a) at the top boundary, high density gradients are observed for the top boundary relative to the bottom one.  For FCC case with $D= 0.5$, moderate stratification of density is seen across the bulk. The behavior near the top  boundary is similar to the SAC case with positive density gradient    due to moderate $\epsilon= 0.45$. However, it is not high enough to observe departures from the adiabatic profile near the bottom boundary.

\subsection{Local thermal boundary layer thickness.}
The strong asymmetry between the boundary layer (BL) dynamics at the top and bottom and the suppression of fluctuations in the SSC case is finally verified by the distribution of the {\em local thermal boundary layer thickness}, which is given in dimensionless and renormalized form by, cf. \cite{Scheel2014} and compare with eq. \eqref{locBL1}, 
%------------------------------------------------------------------------
\begin{equation}
\lambda_{T}^{\rm bot,top}= \Big| \langle T_{\rm sa}(0,1) - T_{\rm sa}(0.5)\rangle_{A,t} \Big | \Bigg| \frac{\partial T_{\rm sa}}{\partial z}\Bigg|^{-1}_{z/H=0,1}\,.
\label{locBL}
\end{equation}          
%------------------------------------------------------------------------
Figure \ref{Fig4} shows the probability density functions (PDFs) of the local thickness evaluated for the four cases at $Ra\simeq 10^6$. It is seen that the distributions at the bottom plane collapse fairly well in the core. Extended left tails are observed for FCC and SSC which also generate the thinnest local BL thicknesses. At the top, the differences between the runs are more significant. While both distributions agree well for the OB-like case (and thus the bottom PDF is shown only), those of the other cases deviate increasingly, most strongly for SSC. The latter distribution is characterized by a small variance and peaked at large local thickness, see also Fig. \ref{Fig1}(a). Quantitatively, the discrepancy between the PDFs at top and bottom can be determined by the Kullback-Leibler (KL) divergence \cite{Goodfellow2016}, which is given by 
%------------------------------------------------------------------------
\begin{equation}
{\rm KL}\left[p^{\rm top}\parallel p^{\rm bot}\right]=\sum_{i=1}^{N_p} p_i^{\rm top}\log\left(\frac{p_i^{\rm top}}{p^{\rm bot}_i}\right)\,,
\label{KL}
\end{equation}          
%------------------------------------------------------------------------
with $p^{\rm top}=p(\lambda_T^{\rm top})$ and $p^{\rm bot}=p(\lambda_T^{\rm bot})$. The index $i$ runs over the bins of both PDFs, $N_p$ is the total number of bins. The inset of panel (b) of the figure shows the KL divergence and detects an increasing deviation of both PDFs which is strongest for the SSC case.

\section{Compressibility in the Mach number--dilatation parameter plane}

In compressible turbulence, it is well known that the system cannot be characterized by the turbulent Mach number alone, but also depends on the  external driving conditions. Recently, Donzis and Panickacheril John \cite{Donzis2020} showed that by including the dilatation parameter $\delta$ along with $M_{t}$, for a variety of driving conditions, at least for homogeneous compressible turbulence, universality for many statistical properties can be demonstrated in terms of these two parameters. Thus the behavior of compressible turbulence depends crucially on where the system is in the $M_{t}$--$\delta$ phase plane.

Even though the extension of the ideas from ref. \cite{Donzis2020} to inhomogenous flows is not fully understood nor studied, recent investigations \cite{Baranwal2022, BDB2023} demonstrate that the Mach number along with the dilatation parameter is again required to characterize heat flux at the boundaries in compressible turbulent channel flows for different thermal boundary conditions. We stress that we assume  $\langle \epsilon_{d} \rangle/\langle \epsilon_{s} \rangle \propto \delta^{2} $ with a proportionality constant ${\cal O}(1)$ to arrive at the definition of $\delta$ in \eqref{eq:deltadef}. Consequently, we only get a relative estimate of the compressibility conditions across the layer depth for the 4 different cases.  Thus comparing our data in the $M_{t}$--$\delta$ phase plane will give us a first-order estimate of the compressibility conditions in the convection flows. 

%----------------------------------------------------------
\begin{figure}[t]
\centering
\includegraphics[width=1.0\linewidth]{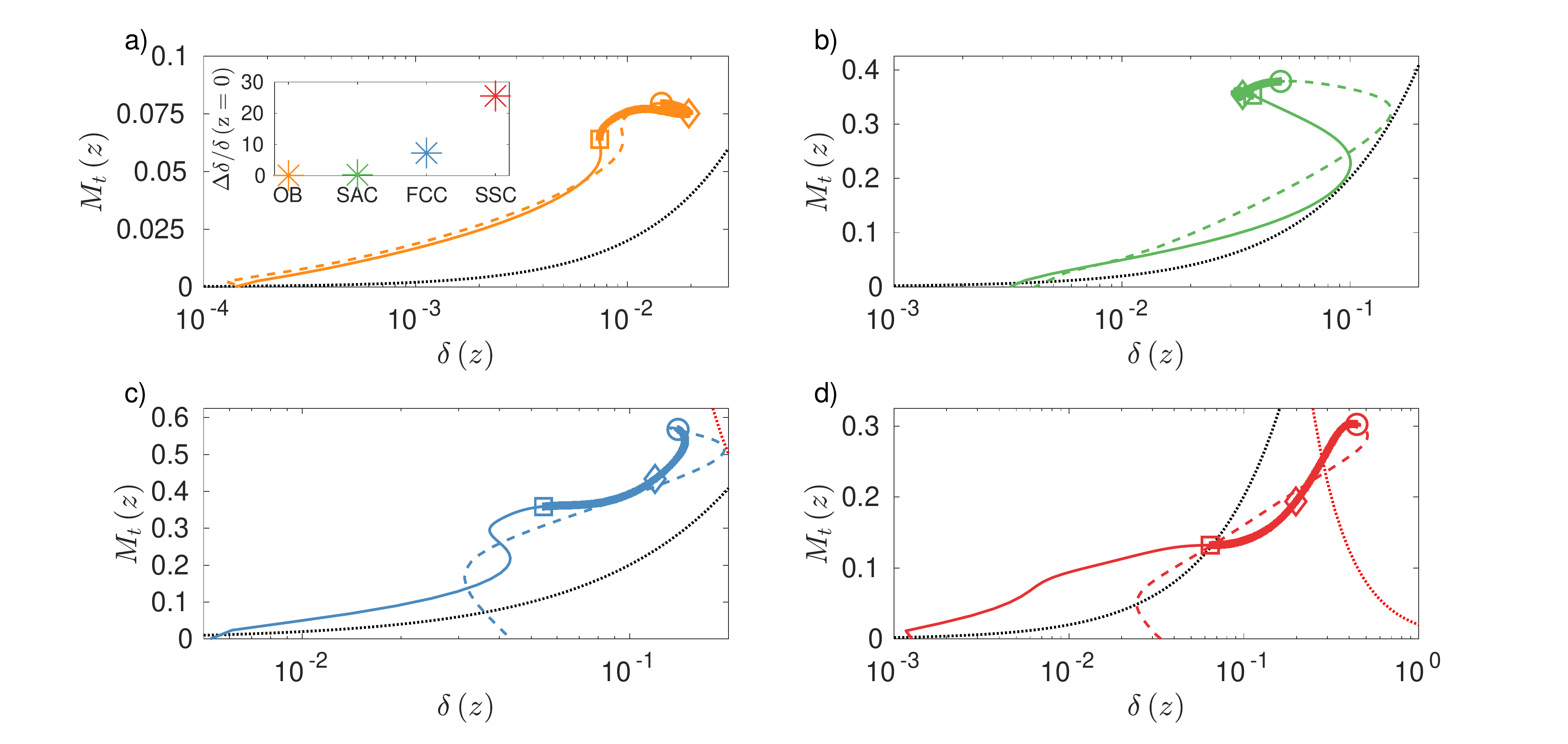}
\caption{Compressible convection regimes in $M_{t}$--$\delta$ phase plane for $Ra \approx 10^6$ (runs 5-8). The data for $M_{t}(z)$ and $\delta(z)$ are taken from Figs. \ref{Fig2}(b) and (c), respectively. (a) case OB, (b) case SAC, (c) case FCC, and (d) case SSC. The hollow circles and squares for all cases correspond to the location of maximum $M_{t}$ near the top and bottom  boundaries. The diamonds correspond to the layer center. The region between the square and circle is considered as the bulk and marked by crosses. The top boundary layer region between top wall and the circle is shown as a dashed line in each panel. Similarly the solid line corresponds to the bottom boundary layer region between the bottom  boundary and the square. The dotted black and red lines in all figures correspond to the equipartition line $f_{\rm eq}=\delta \sqrt{\delta^2 + 1}/M_{t} = 0.5$ and a skewness line $\left(\delta^{2} M_{t} = 0.025 \right)$, respectively \cite{Donzis2020}.  The inset of (a) shows the ratio $\Delta \delta / \delta(z=0)$ near the regions between the top and bottom walls for the four cases.}
\label{Fig6}
\end{figure}
%----------------------------------------------------------

To this end, we combine the data from Figs. \ref{Fig2}(b)  and  (c) in the new Fig. \ref{Fig6} to represent the variation of $\delta(z)$ and $M_{t}(z)$ for all 4 cases in the phase plane. In all panels, the circles and squares correspond to the location of the maximum $M_{t}$ near the top and bottom boundary, respectively. These locations are considered to be the edge of the viscous or thermal boundary layer.  The diamonds stand for the midsection of the layer.  The solid and dashed lines correspond to the points inside the bottom and top boundary respectively. The bulk region is considered to be between the circles and squares. 

We see that $M_{t} \approx 0$ very close to the wall for all cases.  However, here we find that except for the OB case in (a),  $\delta$ has a finite value near the boundaries. The black-dotted  line in all the figures is the equipartition line, $f_{\rm eq}= \delta \sqrt{\delta^{2} + 1}/M_{t}= 0.5$ from Donzis and Panickacheril John \cite{Donzis2020}. This  line demarcates two different regimes of compressible turbulence in homogeneous isotropic turbulence. In the $f_{\rm eq} < 0.5$,   the solenoidal or incompressible nature of pressure dominates over the acoustic or dilatational nature. For $f_{\rm eq} > 0.5$, the acoustic pressure starts to dominate over the solenoidal pressure and it is called weak equipartition regime. 

From Fig. \ref{Fig6}(a), one observes that  for the OB case, both the top and bottom boundaries almost overlap each other perfectly in the phase plane implying a nearly perfect top-down symmetry of both boundary layer. For the SAC case (b), very close to the wall, unlike the OB case, the top and bottom boundaries do not overlap, but are close to each other. This again is consistent with the slight asymmetry seen for the SAC case in Figs. \ref{Fig4}. Similar to  Fig. \ref{Fig4}, an increase of $D$ is connected with increasing differences in the phase plane, as visible for the cases FCC and SSC in panels (c) and (d), respectively. For the SSC case, very near the wall the difference between the curves for the top and bottom boundary are strongest. 

Moreover, we notice that if the estimate of  $f_{\rm eq}$ for homogeneous flows is also valid for the present inhomogeneous system, then the top and bottom boundary layers of SSC would be found in different regimes of compressible turbulence.  One can quantify the difference with the measure, $\Delta \delta/\delta (z= 0)$, where $\Delta \delta = \delta (z= H) - \delta (z= 0)$. We plot this measure as an inset in panel (a) for all 4 cases. A striking similarity between $\Delta \delta/\delta \left(z= 0\right)$ and the KL divergence is found. Clearly this indicates that the relative strength of the dilatational motions plays an important role in the boundary layer dynamics of compressible convection. Finally, we notice for the low-$D$ cases SAC and OB, that the variability of compressibility conditions in the phase plane is very limited implying homogeneity  across the bulk. In fact, for the SAC, this range is the most narrow one. The opposite trend is observed as the dissipation number $D$ increases which indicates an inefficiency of the turbulent mixing. 

\section{Detailed Energy Budget Analysis}

The energy balance equation \eqref{eq:ener} can also be written as 
\begin{subequations}
    \begin{equation}
C_{p} \frac{\partial \left(\rho T\right)}{\partial t} + C_{p} \frac{\partial \left(\rho  u_{j} T\right)}{\partial x_{j}}  - \frac{\partial p}{\partial t}  - u_{j} \frac{\partial p }{\partial x_j} = 
\frac{\partial}{\partial x_{j}}\left(k \frac{\partial T}{\partial x_{j}}\right)
+ \sigma_{ij} S_{ij} 
\label{eq:ener1a}
\end{equation}
Averaging over horizontal planes and time, along with the assumption of a statistically steady state, we get 
\begin{equation}
C_{p} \frac{  \partial \langle \rho  u_{j} T \rangle_{A,t} }{\partial x_{j}}  -  \Big\langle u_{j} \frac{\partial p }{\partial x_{j}} \Big\rangle_{A,t}  = 
\frac{\partial}{\partial x_{j}}\Big\langle k \frac{\partial T}{\partial x_{j}}\Big\rangle_{A,t} 
+ \langle \sigma_{ij} S_{ij} \rangle_{A,t} 
\label{eq:ener1b}
\end{equation}
Since we have homogeneity in both horizontal directions, the equation can further be simplified to 
\begin{equation}
C_{p} \frac{  d \langle \rho  u_z T \rangle_{A,t} }{dz }  - \Big\langle u_{j} \frac{\partial p }{\partial x_{j}} \Big\rangle_{A,t}  = 
\frac{d}{dz}\Big\langle k \frac{d T_{\rm sa}}{d z}\Big \rangle_{A,t} 
+ \langle \sigma_{ij} S_{ij} \rangle_{A,t} 
\label{eq:ener1c}
\end{equation}
Note that we have also replaced $T$ with the superadiabatic temperature, $T_{\rm sa}$ since the adiabatic temperature profile, $\overline{T}(z)=T_{\rm bot}(1-Dz)$, is a linear function of $z$.  Finally integrating from the bottom wall to an arbitrary $z$, we get   
\begin{equation}
C_{p}  \langle \rho u_z T \rangle_{A,t}(z)  
-\Big\langle k \frac{d T_{sa}}{d z}\Big\rangle_{A,t} (z)
- \int_{0}^{z} \Big\langle u_{j} \frac{\partial p }{\partial x_{j}} \Big\rangle_{A,t}  dz' 
- \int_{0}^{z} \langle \sigma_{ij} S_{ij} \rangle_{A,t} \left(z' \right) dz' = \text{const.} 
\label{eq:flux}
\end{equation}
\end{subequations}
For the OB Rayleigh-B\'{e}nard convection case, the contribution from the last two terms constituting the compression work and energy dissipation, respectively, are negligible.  After normalizing with $k_0\Delta T/(C_p \rho_{\rm ref}H)$, one would get the standard Nusselt number definition \cite{Chilla2012}, 
\begin{equation}
 Nu(z)=\dfrac{\langle u_z T \rangle_{A,t}(z)  - \dfrac{k_0}{C_p\rho_{\rm ref}} \Big\langle  \dfrac{d T_{sa}}{d z}\Big\rangle_{A,t}(z)}{\dfrac{k_0}{C_p\rho_{\rm ref}}\dfrac{\Delta T}{H}}\,, 
\label{eq:incompflux}
\end{equation}
which has to be constant for each horizontal  plane $0\le z\le H$. Normalizing eq. \eqref{eq:flux} with  $\epsilon T_{\rm bot} k/H$, one gets 
\begin{align}
\underbrace{
\left[\frac{C_{p} H}{k\epsilon T_{\rm bot}}\right] \langle \rho u_zT \rangle_{A,t} }_{=J_{c}(z)}     
\underbrace{
-\left[\frac{H}{k\epsilon T_{\rm bot}}\right]  \int_{0}^{z} \Big\langle u_{j} \frac{\partial p }{\partial x_{j}} \Big\rangle_{A,t} dz'
}_{=J_{p}(z)}
\underbrace{ 
- \left[\frac{ H}{\epsilon T_{\rm bot}}\right]
\Big\langle  \frac{d T_{\rm sa}}{d z}\Big\rangle_{A,t}}_{=J_{d}(z)}
\underbrace{
-\left[\frac{H}{k\epsilon T_{\rm bot}}\right]  \int_{0}^{z} \langle \sigma_{ij} S_{ij} \rangle_{A,t} dz'
}_{=J_{\langle \epsilon\rangle}(z)}=\text{const.} 
\label{eq:compflux}
\end{align}
Equation \eqref{eq:compflux} contains the diffusive current $J_{d}(z)$ at the boundaries $z = 0$ and $H$ only; the other terms vanish due to the no-slip boundary conditions for the velocity field. This confirms our definition of the Nusselt number in \eqref{eq:nuss}. In eq. \eqref{eq:compflux}, the terms $J_{c}(z)$, $J_{p}(z)$, $J_{d}(z)$, and  $J_{\langle \epsilon \rangle}(z)$ denote the  convective heat current, the current due to compression work, the diffusive heat current, and energy dissipation current, respectively.
%------------------------------------------------------
\begin{figure}[t]
\centering
\includegraphics[width=0.9\linewidth]{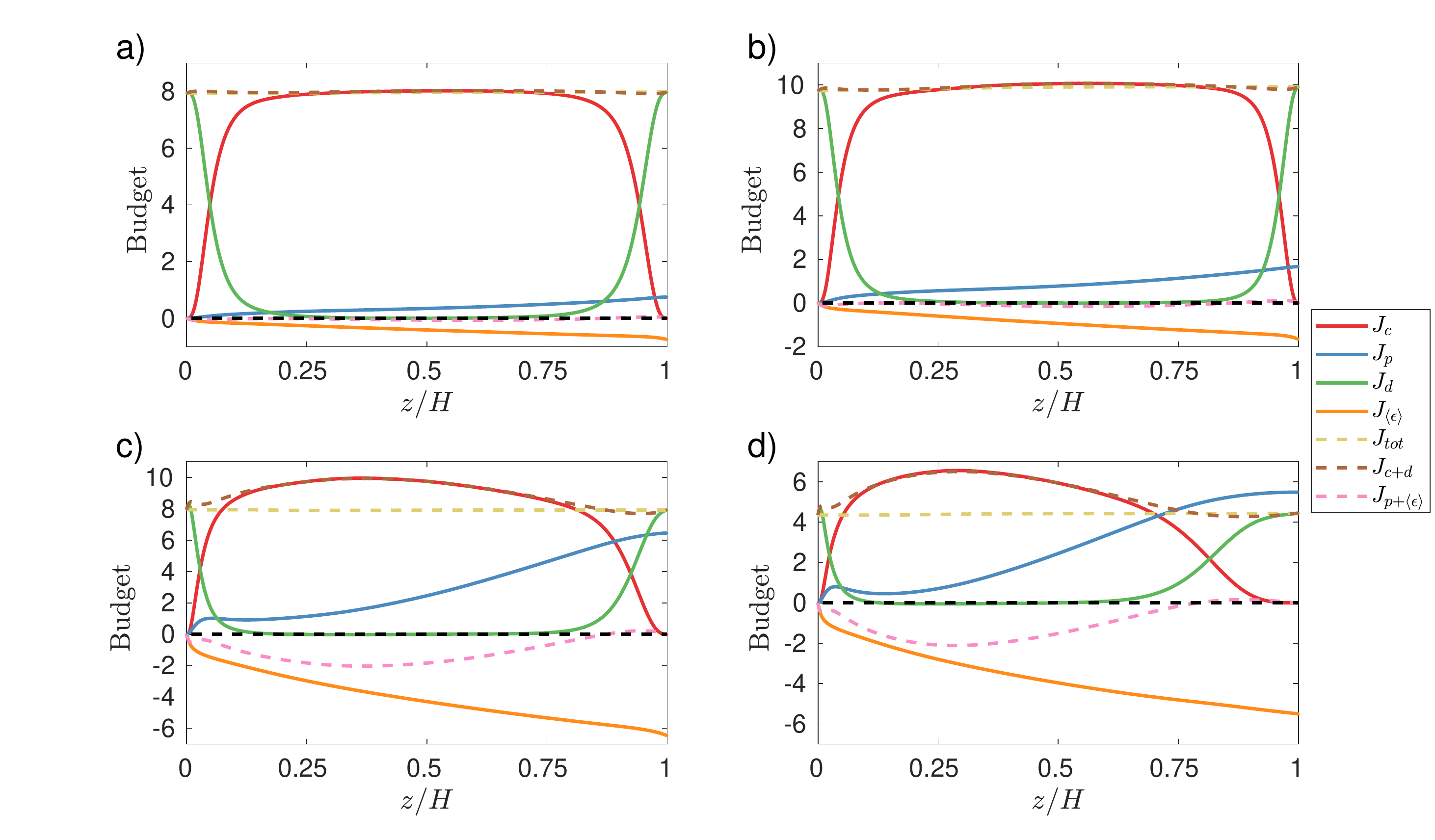}
\caption{Detailed energy budget versus layer height for all 4 convection cases. (a) case OB, (b) case SAC, (c) case FCC, and (d) and SSC. The definitions of the fluxes: $J_{c}$, $J_{d}$, $J_{p}$ and $J_{\langle\epsilon \rangle}$ are given in Sec. VII. We also show $J_{\rm tot}= J_{c} + J_{p} + J_{d} +J_{\langle\epsilon \rangle} $, and the partial sums $J_{c+d}= J_{c} + J_{d}$, and $J_{p + \langle \epsilon \rangle}=  J_{p}  +J_{\langle\epsilon \rangle}$. All data are again for $Ra=10^6$. Note that in panels (a) and (b) the flux $J_{\rm tot}$ practically coincides with $J_{c+d}$.}
\label{Fig7}
\end{figure}
%------------------------------------------------------
We plot these fluxes and their partial and total sums as function of depth $z$ for all 4 convection cases in  Fig. \ref{Fig7} at $Ra=10^6$. The sum of all these terms is a constant, i.e.,
\begin{equation}
Nu(z) = J_{\rm tot}(z) = J_{c}(z) + J_{p}(z) + J_{d}(z) + J_{\langle \epsilon \rangle}(z) = Nu\,. 
\label{eq:totsum}
\end{equation}
See again table \ref{tab:simusupp} for the Nusselt number of all cases. 
 
In panels (a) and (b) of Fig. \ref{Fig7}, the compression work and dissipation terms perfectly cancel each other over the entire domain, thus the relation has to be very close to the exact Rayleigh-B\'{e}nard case, $J_{c} + J_{d} = Nu$. For the SAC case, due to high superadiabaticity with $\epsilon= 0.8$, the magnitude of these two canceling terms increases with height.   

For the cases FCC and SSC in panels (c) and (d) significant differences are observed. The sum $J_{p}(z) + J_{\langle\epsilon\rangle}(z) \ne 0$ across the domain. Also, the magnitude of both terms is significantly higher. Since the diffusive flux $J_{d}\approx 0$ in the bulk, the difference between the $J_{p}$ and $J_{d}$ has to be compensated by $J_{c}$. Thus unlike the incompressible Rayleigh-B\'{e}nard case, $J_{c} \ge Nu$ rather than $J_{c} = Nu$ in the bulk. In other words, a fraction of the convective current in the bulk is used for balancing between the compression work and dissipation contributions, which might explain the decreased amount of transported heat for the higher $D$ cases.

\section{ Final discussion. \label{sec8}} 
The central objective of this work was a systematic exploration of  different regimes of fully compressible convection which are determined by the superadiabaticity $\epsilon$ and the dissipation number $D$. For sufficiently large amplitudes of both parameters, convection proceeds beyond the Oberbeck-Boussinesq and anelastic limits. We focused on genuine compressibility effects and excluded other non-Oberbeck-Boussinesq contributions which can arise from complicated dependencies of the material parameters, such as of the dynamic viscosity or the thermal conductivity on the thermodynamic state variables $T$ and $p$. We identified three limiting cases in the three corners of the triangular parameter plane which is spanned by superadiabaticity and dissipation number. These are the Oberbeck-Boussinnesq-like limit for $\epsilon\ll 1$ and $D\ll 1$, the strongly superadiabtic convection case for $\epsilon\to 1$ and $D\ll 1$, and the strongly stratified convection case for $\epsilon\ll 1$ and $D\to 1$. A fourth regime maximizes the Mach number and can be considered as a blend of the latter two regimes with $\epsilon\sim D\sim 0.5$.

We have systematically explored the mean profiles of central quantities, such as the temperature, the velocity fluctuations, or the convective heat flux. A further aspect of the present study was to transfer the analysis concepts from homogeneous, isotropic compressible turbulence to the present flow with one inhomogeneous spatial direction, e.g. by monitoring the turbulent Mach number and dilatation parameter in the phase plane, see Fig. \ref{Fig6}. 

The majority of convection problems are discussed either in the Oberbeck-Boussinesq or the anelastic approximations. These approximations can be well described in terms of the  asymptotic limits of $\epsilon$ and $D$. The  Oberbeck-Boussinesq approximation corresponds to $\epsilon \to 0$ and subsequently $D \to 0$. The anelastic approximation stands $\epsilon \to 0 $ and a finite $D$, such that the Mach number, $M_f \rightarrow 0$. Other low-Mach-number approximations \cite{durran1989,klein2012,almgren2006}  have been proposed in the literature. These approximations are less restrictive compared to the anelastic one. Unlike the AE approximation, these models allow for large variations of density and temperature about their mean values. However, pressure fluctuations are neglected.  Where would these low-Mach-number approximations be found in the $\epsilon-D$ phase plane? One the one hand, they should be valid in the anleastic limit, on the other hand in the limit of vanishing $D$ with a finite superadiabaticity $\epsilon$.  The finite $\epsilon$ allows for large variations of density and temperature. However, $D \to 0$, and thus $M_f \rightarrow 0$,  implies that that the pressure variations are negligible. All our simulation have a considerable Mach number near the top boundary; thus these approximations are not valid.

Of particular interest from our point of view is the SSC case. Highly top-down-asymmetric compressible convection is obtained for the strong stratification case, i.e., for $D\to 1-\epsilon$. In this regime, we detect a sublayer with strongly reduced velocity, temperature fluctuations and a reduced convective heat flux. These properties are also documented by the resulting Nusselt and Reynolds numbers in table \ref{tab:simusupp}. Out of this layer self-focusing thermal plumes are ejected deep into the bulk of the strongly stratified convection zone. A focusing effect in compressible turbulence has been recently reported in laboratory experiments by Manzano et al. \cite{Manzano2022}. 

Even though our compressible convection simulations, which do not include magnetic fields and radiation transfer, are certainly much simpler than stellar convection flows, they might provide interesting new insights on how the heat is transported by nonlocal coherent downwelling plumes that detach from the top boundary layer. As suggested in the review of Spruit \cite{Spruit1997}, this might be particularly important when strong density stratification exists and the level of turbulence is strongly reduced in regions between the downwelling plumes. The latter point might be connected to the solar conundrum in the upper solar convection zone which reports anomalously weak velocity fluctuations for depths $z\gtrsim 0.92 R_{\odot}$, determined by helioseismology \cite{Hanasoge2012}. We thus think that our present DNS, in particular those in the SSC regime, might provide an interesting testing bed for the exploration of some basic processes and concepts in such flows, e.g. the formation of these plumes and their turbulence production by shear instabilities. Indeed estimated dissipation numbers $D$ would be significant when using data from model S of the solar interior \cite{Christensen1996,Schumacher2020}: 1) If we take the surface layer with a depth of 0.01 $R_{\odot}$, which corresponds to $H\approx 6900$ km, we have $T_{\rm bot} \simeq 4.6\times 10^4$ K, $g\simeq 290$ m/s$^2$, $c_p\simeq 4.7 \times 10^4$ J/(kg K). This results to $D\approx 0.93$. 2) If we would take the lower half of the convection zone as a convection layer which corresponds to $H\approx 10^5$ km, we have $T_{\rm bot} \simeq 2.2\times 10^6$ K, $g\simeq 430$ m/s$^2$, $c_p\simeq 3.5 \times 10^4$ J/(kg K). This results to a much smaller dissipation number of $D\approx 0.56$. Note that the values for $g$ and $c_p$ have been estimated as geometric means with three reference points for this coarse estimate. 

Motivated by this discussion, we will continue to explore the highly stratified  regime at higher Rayleigh and lower Prandtl numbers in the future. \\

{\em Acknowledgements.} 
The work of J.P.J. is supported by grant no. SCHU 1410/31-1 of the Deutsche Forschungsgemeinschaft and a Postdoctoral Fellowship of the Alexander von Humboldt Foundation. The authors gratefully acknowledge the Gauss Centre for Supercomputing e.V. (https:// www.gauss-centre.eu) for funding this project by providing computing time through the John von Neumann Institute for Computing (NIC) on the GCS Supercomputer JUWELS at J\"ulich Supercomputing Centre (JSC). We are thankful to Diego A. Donzis and Akansha Baranwal for sharing the cDNS code which has been adapted in this work. Finally, we thank Janet D. Scheel for discussions.

\end{document}